\documentclass[%
reprint,
superscriptaddress,
 amsmath,amssymb,
 aps,
 prb,
]{revtex4-2}

\usepackage[colorlinks, linkcolor=blue, anchorcolor=blue, citecolor=blue, urlcolor=blue]{hyperref} 
\usepackage{graphicx}
\usepackage{bm}
\usepackage{float}
\usepackage{multirow}
\usepackage{setspace}

\begin{document}

\title{Strongly Enhanced Charge-Density Waves and Correlated Insulating State in Atomically Thin 1$T$-TaS$_2$}

\author{Gan~Liu}
\thanks{These authors contributed equally to this work}
\author{Yulu~Liu}
\thanks{These authors contributed equally to this work}
\author{Qiling~Luo}
\thanks{These authors contributed equally to this work}
\affiliation{National Laboratory of Solid State Microstructures and Department of Physics, Nanjing University, Nanjing 210093, China}

\author{Zhentao Huang}
\affiliation{National Laboratory of Solid State Microstructures and Department of Physics, Nanjing University, Nanjing 210093, China}
\affiliation{Institute of High Energy Physics, Chinese Academy of Sciences, Beijing 100049, China}
\affiliation{Spallation Neutron Source Science Center, Dongguan 523803, China}

\author{Kenji~Watanabe}
\affiliation{Research Center for Functional Materials, National Institute for Materials Science, 1-1 Namiki, Tsukuba 305-0044, Japan}

\author{Takashi~Taniguchi}
\affiliation{International Center for Materials Nanoarchitectonics, National Institute for Materials Science,  1-1 Namiki, Tsukuba 305-0044, Japan}

\author{Meiyu~Wang}
\affiliation{Center for Shared Scientific Research Facilities, Nanjing University, Nanjing 210093, China}

\author{Jinsheng~Wen}
\author{Yi~Lu}
\email{yilu@nju.edu.cn}
\author{Xiaoxiang~Xi}
\email{xxi@nju.edu.cn}
\affiliation{National Laboratory of Solid State Microstructures and Department of Physics, Nanjing University, Nanjing 210093, China}
\affiliation{Collaborative Innovation Center of Advanced Microstructures, Nanjing University, Nanjing 210093, China}
\affiliation{Jiangsu Physical Science Research Center, Nanjing 210093, China}

\begin{abstract}
We investigate thickness-dependent charge-density-wave (CDW) transitions in 1$T$-TaS$_2$ using temperature-dependent Raman spectroscopy and electrical transport. Raman measurements show that the incommensurate, nearly commensurate, and commensurate CDW phases persist down to the monolayer limit. As the thickness is reduced, the transition temperatures increase, accompanied by an orders-of-magnitude rise in sheet resistance and a sharp reduction in the carrier localization length. The first-order hysteretic CCDW-NCCDW transition is uniquely absent in the monolayer. Calculations suggest that the enhanced CDW in thin layers originates from strengthened Coulomb interactions due to reduced out-of-plane screening, particularly in the nonlocal component. These findings highlight the cooperative roles of electron correlation, electron-phonon interaction, and interlayer coupling in shaping the ground state and transition dynamics of atomically thin 1$T$-TaS$_2$, opening pathways for engineering correlated phases in two-dimensional CDW systems.
\end{abstract}

\maketitle

The interplay among lattice, charge, orbital, and spin degrees of freedom in solids gives rise to diverse electronic properties. A well-known example is the charge-density wave (CDW), a broken-symmetry state in which electronic density modulations are coupled to lattice distortions~\cite{Gruner2000}. The layered transition-metal dichalcogenide 1$T$-TaS$_2$ epitomizes this physics, showing metal-insulator transitions driven by CDW formation~\cite{Wilson_1975}, superconductivity under pressure or doping~\cite{Sipos_2008,Ang_2012,Yu2015,Dong_2021}, and a possible quantum spin-liquid state~\cite{Law2017,Martin_2017,He_2018}. These phenomena make 1$T$-TaS$_2$ a valuable platform for exploring the coexistence and competition of distinct quantum phases.

Bulk 1$T$-TaS$_2$ exhibits a cascade of CDW transitions upon cooling, entering an incommensurate phase (ICCDW) near 550~K, a nearly commensurate phase (NCCDW) around 350~K, and a fully commensurate phase (CCDW) below 180~K~\cite{Sipos_2008}. The microscopic origins of these phases remain unsettled. Early studies attributed the formation of the ICCDW phase to Fermi-surface nesting~\cite{Woolley_1977,PMyron_1975,Bovet_2004,Clerc_2006}, whereas more recent work highlights the role of electron-phonon coupling~\cite{Liu_2009}. The NCCDW and CCDW phases involve further complexity due to possible electron correlation effects~\cite{Clerc_2006,Rossnagel_2011,Zhang_2014,Liu_2009,Cho_2015,Yi_2018}. Whether the insulating CCDW state should be ascribed to the opening of a Mott gap stabilized by lattice reconstruction~\cite{Fazekas_1980} or to a single-particle gap induced by interlayer Peierls dimerization~\cite{Ritschel_2015,Lee_2019} remains debated.

Recent advances in exfoliation and thin-film growth have opened new opportunities to address these questions in the two-dimensional (2D) limit. Because electron correlation, electron-phonon coupling, and interlayer dimerization all depend on dimensionality, atomically thin samples provide a useful setting to disentangle their roles. Previous transport studies on exfoliated 1$T$-TaS$_2$ flakes reported that the NCCDW and CCDW phases vanish below critical thicknesses of 3~nm and 10~nm, respectively~\cite{Yu2015}, whereas Raman scattering studies suggested that the CCDW phase persists even in the monolayer limit~\cite{Albertini_2016,He2016}. Molecular beam epitaxy (MBE) has further enabled the growth of monolayer 1$T$-TaS$_2$ films that exhibit the CCDW phase~\cite{Lin_2020} and show evidence of a quantum spin liquid state~\cite{Chen_2025}. However, a consistent picture of how the series of CDW phases and their associated electronic states evolve with reduced dimensionality remains elusive.

\begin{figure}[t]
\centering
\includegraphics[width=\linewidth]{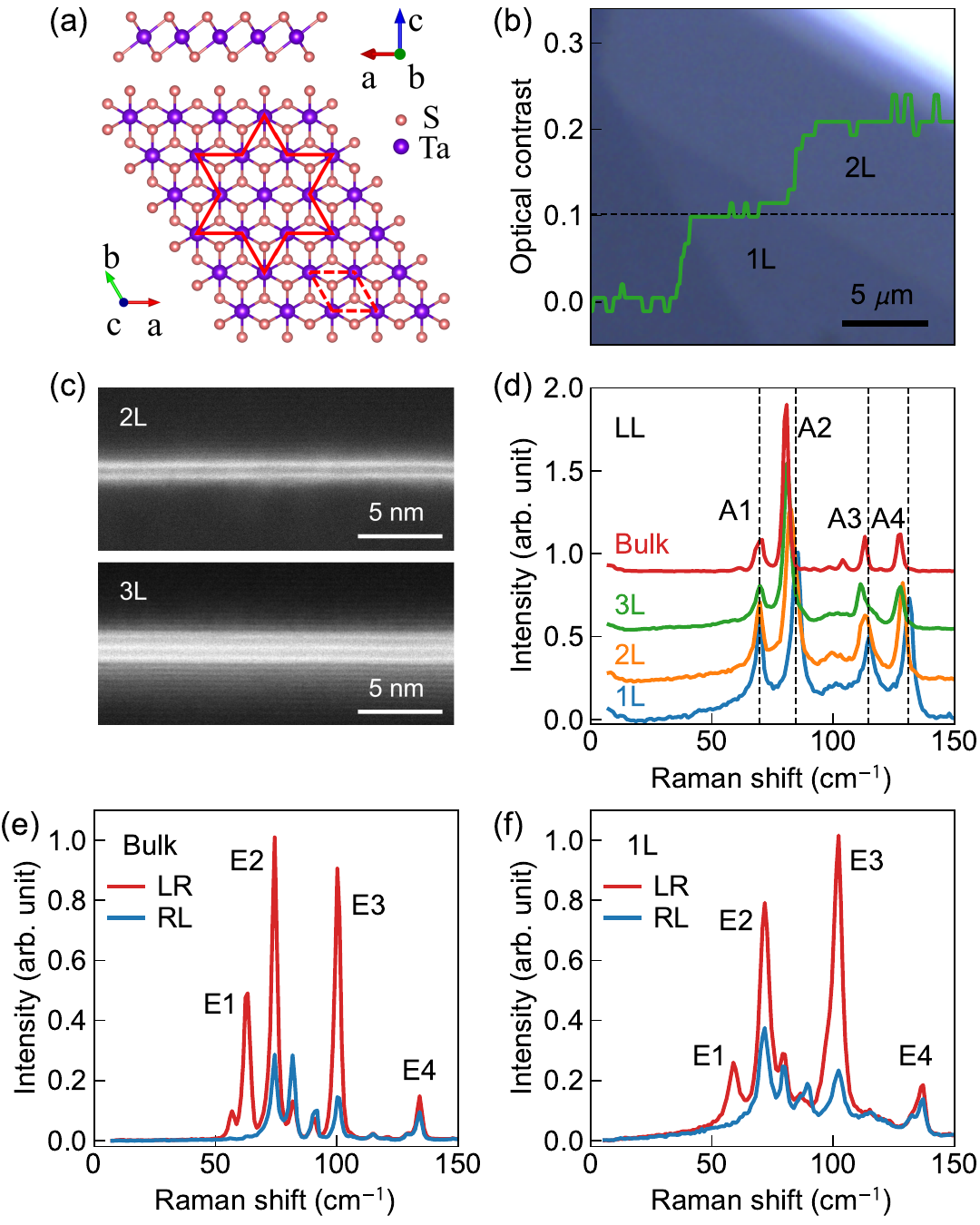}
\caption{(a) Crystal structure of monolayer 1$T$-TaS$_2$. The dotted lines mark the unit cell of the undistorted $1T$ structure. The hexagram highlights the SD pattern in the CCDW phase. (b) Optical micrograph of exfoliated flakes on PDMS, with the solid line showing the red-channel optical contrast along the dotted line. (c) Cross-sectional STEM images showing bilayer (upper panel) and trilayer (lower panel) 1$T$-TaS$_2$ encapsulated by h-BN. (d) Thickness-dependent Raman spectra in the LL configuration at 10 K, vertically offset for clarity. The main $A_g$ phonons (A1--A4) are indicated, with the monolayer frequencies marked by dashed lines. (e)--(f) Raman spectra in the LR and RL configurations for bulk (e) and monolayer (f) samples at 10 K. All spectra in (d)--(f) are normalized to the most intense peak.}
\label{Fig1}
\end{figure}

In this Letter, we report a comprehensive thickness-dependent study of CDW transitions in 1$T$-TaS$_2$. Temperature-dependent Raman spectroscopy tracks the phonon evolution across all CDW phases and confirms that the CCDW phase persists in the monolayer limit. Electrical transport reveals extremely high sheet resistance and strong carrier localization in monolayer and bilayer crystals. The resulting temperature-thickness phase diagram shows a pronounced enhancement of both CCDW-NCCDW and NCCDW-ICCDW transition temperatures with decreasing thickness. Theoretical calculations show that nonlocal Coulomb interactions are essential to reproduce the thickness dependence of phonon frequencies and simultaneously strengthen the CDW lattice distortion. These results support the correlated nature of the insulating state in atomically thin 1$T$-TaS$_2$.

Monolayer 1$T$-TaS$_2$ consists of a central Ta layer octahedrally coordinated by S atoms [Fig.~\ref{Fig1}(a)]. Bulk crystals are formed by stacking such layers through van der Waals coupling, allowing the isolation of atomically thin flakes by mechanical exfoliation. Figure 1(b) shows a typical optical micrograph of atomically thin 1$T$-TaS$_2$ on a polydimethylsiloxane (PDMS) substrate. The optical contrast, defined as $\alpha = I_{\mathrm{F}}/I_{\mathrm{Sub}}-1$, scales linearly with layer number, enabling identification of monolayer (1L) and bilayer (2L) regions. Here $I_{\mathrm{F}}$ and $I_{\mathrm{Sub}}$ are the flake and substrate intensities, respectively. These thickness assignments were further confirmed by cross-sectional scanning transmission electron microscopy (STEM) and atomic force microscopy [Fig.~\ref{Fig1}(c) and Supplemental Material, Note~1~\footnote{See Supplemental Material for experimental methods (sample preparation and device fabrication, thickness determination, transport and optical measurements), Raman mode fitting analysis, additional Raman data, analysis of sheet resistance and critical temperatures, and additional calculation results.}], which also yield an interlayer spacing of 6--7~{\AA}~\cite{Mattheiss_1973}. 

In bulk 1$T$-TaS$_2$, the hallmark of the CCDW phase is the star-of-David (SD) cluster [Fig.~\ref{Fig1}(a)]. Their close packing in the $ab$-plane produces a $\sqrt{13}\times\sqrt{13}R13.9^{\circ}$ reconstructed superlattice, which enlarges the unit cell and rotates the crystal axis~\cite{Wilson_1975}. This reconstruction folds the Brillouin zone and activates new Raman modes that provide clear phononic signatures of the CDW. The $E_g$ and $A_g$ modes of the resulting $C_{3i}$ point group can be distinguished using helicity-resolved Raman scattering~\cite{Yang2022,Liu_2023}. In this scheme, the equivalent LL and RR polarization configurations probe the $A_g$ modes, where L and R denote left- and right-circular polarization, respectively. Thickness-dependent Raman spectra acquired at 10~K in the LL channel are shown in Fig.~\ref{Fig1}(d), featuring intense zone-folded modes associated with Ta-dominated acoustic phonons below 150~cm$^{-1}$~\cite{Mijin_2021}. S-dominated modes lying above 150~cm$^{-1}$ are much weaker~\cite{Yang2022,Liu_2023} and therefore not discussed here. For samples of all thicknesses, the spectra show four intense modes, A1--A4, indicating that the CCDW phase persists down to the monolayer limit. This observation agrees with the work by Albertini \textit{et al.}~\cite{Albertini_2016} but contrasts with reports of its absence in few-layer and monolayer samples~\cite{Ramos_2019,Sanders_2016}. The discrepancy may arise from improved quality of our samples, enabled by preparation in an inert-gas atmosphere and h-BN encapsulation, as reflected by the sharp phonon modes even in the monolayer. 

The $E_g$ modes are characteristic of the ferro-rotational CCDW phase, which hosts two mirror-symmetric domain states with the superlattice rotated $\pm 13.9^{\circ}$ relative to the pristine lattice~\cite{Yang2022,Liu_2023}. In this phase, Raman spectra in the LR and RL channels become nonequivalent, reflecting nonzero off-diagonal components of the Raman tensor~\cite{Yang2022}. Such behavior is observed at all thicknesses [Fig.~\ref{Fig1}(e)--\ref{Fig1}(f)]. In addition, polarization dependence of optical second harmonic generation for the monolayer sample reveals its threefold rotational symmetry (Supplemental Material, Fig.~2), a signature of the electric quadrupolar order associated with the ferro-rotational CDW~\cite{Luo_2021}. 

\begin{figure}[t]
\centering
\includegraphics[width=\linewidth]{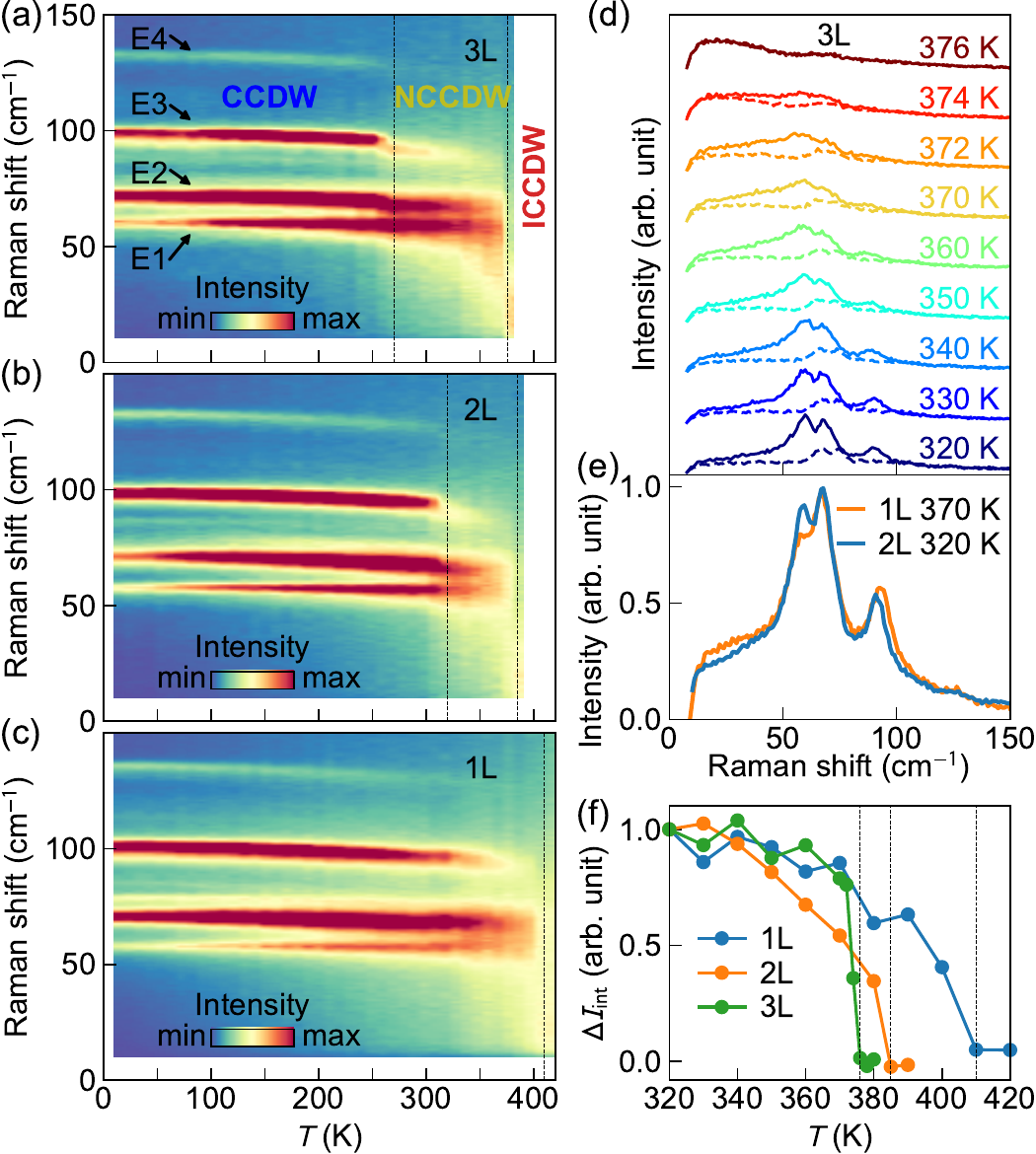}
\caption{(a--c) Temperature-dependent Raman intensity maps for 1--3 layer samples measured in the LR configuration. (d) Raman spectra of the trilayer sample in the LR (solid) and RL (dotted) channels at selected temperatures. (e) Comparison of Raman spectra for monolayer and bilayer samples. (f) Temperature dependence of the integrated differential spectra, normalized to the 320~K value. Vertical dashed lines indicate transition temperatures.}
\label{Fig2}
\end{figure}

\begin{figure}[t]
\centering
\includegraphics[width=1.0\linewidth]{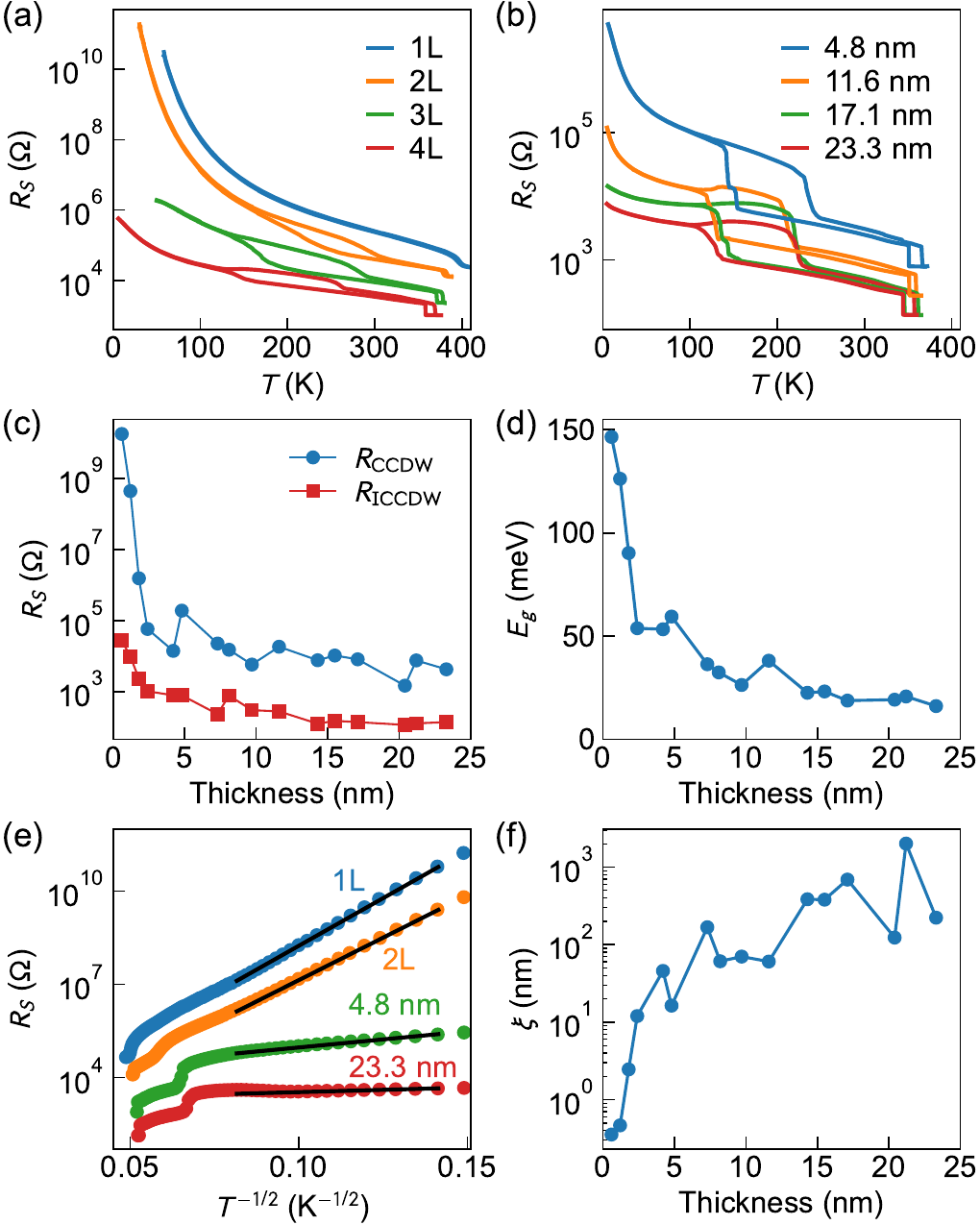}
\caption{(a–b) Temperature-dependent sheet resistance for 1--4 layer samples (a) and thicker samples (b). (c) Thickness-dependent sheet resistance in the CCDW (60~K) and ICCDW phases. (d) Activation gap $E_g$ as a function of thickness. (e) $R_\mathrm{s}$ vs.\ $T^{-1/2}$ (with a logarithmic y-axis) for samples of varying thickness. Symbols are experimental data and solid lines are linear fits. (f) Thickness dependence of the localization length.} 
\label{Fig3}
\end{figure}

We now examine the temperature dependence of the Raman response, which reveals how the CDW transitions evolve with reduced thickness. Figure~\ref{Fig2}(a) presents the temperature-dependent Raman intensity color plot for a trilayer sample measured in the LR channel during warming. Abrupt phonon changes mark the first-order CCDW-NCCDW and NCCDW-ICCDW transitions (see mode analysis in Supplemental Material, Note~2). The disappearance of E1 and E2 modes near 376~K identifies the NCCDW-ICCDW transition, corroborated by the vanishing differential spectra $\Delta I = I_{\mathrm{LR}} - I_{\mathrm{RL}}$, which track the loss of ferro-rotational order from the NCCDW to ICCDW phase [Figs.~\ref{Fig2}(d),~\ref{Fig2}(f)]~\cite{Yang2022,Liu_2023}. Similar transitions are observed in bilayer and thicker samples [Fig.~\ref{Fig2}(b) and Supplemental Material, Note~3], with critical temperatures systematically enhanced as thickness decreases below four layers. In the monolayer, the first-order NCCDW-ICCDW transition persists, with its critical temperature further raised to $\sim$410 K, whereas the first-order CCDW-NCCDW transition disappears [Figs.~\ref{Fig2}(c), \ref{Fig2}(f)]. The monolayer spectrum at 370~K closely resembles the bilayer spectrum at 320~K in the NCCDW phase [Fig.~\ref{Fig2}(e)], indicating a continuous CCDW-NCCDW transition in the monolayer limit.

The thickness dependence of the CDW transitions was further investigated by electrical transport. Figure~\ref{Fig3}(a) shows the temperature-dependent sheet resistance of 1--4 layer samples, all of which exhibit the first-order NCCDW-ICCDW transition upon both warming and cooling. A more detailed view is provided in Supplemental Material, Fig.~10. Compared with thicker samples [Fig.~\ref{Fig3}(b) and Supplemental Material, Note~4], the transitions are broadened, indicating enhanced disorder in atomically thin samples. A wide thermal hysteresis window of $\sim$100~K, associated with the CCDW-NCCDW transition, is present in all samples except the monolayer. This agrees with the continuous CCDW-NCCDW transition observed in monolayer Raman spectra [Fig.~\ref{Fig2}(c)], reflecting the essential role of interlayer coupling in making this transition first-order. Supporting this picture, Monte Carlo simulations for bulk 1$T$-TaS$_2$ attributed the hysteresis to energy barriers between distinct stacking orders~\cite{Lee_2019}, which do not exist in the monolayer. Our results contrast with previous reports that the NCCDW-ICCDW and CCDW-NCCDW transitions vanish below critical thicknesses~\cite{Yu2015,Yoshida_2014,Tsen2015} and that the first-order CCDW-NCCDW transition is absent in few-layer samples~\cite{Constant2021}. Figure~\ref{Fig3}(c) shows the thickness-dependent sheet resistance in the CCDW phase (at 60~K) and ICCDW phase (just above the transition), both of which increase with decreasing thickness. Notably, the sheet resistance in the CCDW phase rises steeply by orders of magnitude for samples thinner than four layers. 

Electrical transport in 1$T$-TaS$_2$ is governed by distinct mechanisms in different temperature regimes. At high temperatures, conduction arises from thermal excitation across the band gap, yielding Arrhenius behavior $\ln R \propto E_g/2k_BT$, where $E_g$ is the activation gap and $k_B$ the Boltzmann constant. Analysis of the resistance data above 150~K shows that $E_g$ increases sharply below four layers, reaching 147~meV in the monolayer [Fig.~\ref{Fig3}(d) and Supplementary Note~4]. This value provides a reasonable estimate of the Mott gap in monolayer 1$T$-TaS$_2$, which has been measured to be $\sim$230~meV by scanning tunneling spectroscopy on MBE-grown films~\cite{Lin_2020} and calculated to be $\sim$200~meV in theoretical studies~\cite{Darancet_2014,Bae2025}, based on the criterion of a vanishing density of states. However,  due to the enhanced interlayer charge transport, this method clearly underestimates the gap in thick samples~\cite{Butler2020,Wu2022}.

At low temperatures, thermal activation is suppressed and conduction proceeds via variable-range hopping (VRH). In the Mott VRH framework~\cite{Mott_1968}, assuming a constant density of states at the Fermi level, $\ln R \varpropto(T_0/T)^{\alpha}$ with $\alpha = 1/4$ and 1/3 for 3D and 2D systems, respectively. $T_0$ is a characteristic temperature. When Coulomb interactions open a gap in the density of states, the Efros-Shklovskii (ES) model applies, with a universal exponent $\alpha = 1/2$~\cite{Efros_1975}. For nanometer-thick 1$T$-TaS$_2$, the low-temperature resistance fits the ES form [Fig.~\ref{Fig3}(e) and Supplemental Material, Note~4], consistent with earlier bulk studies where the effect was linked to crystalline defects~\cite{Inada1983}. From the ES fits, $T_0$ is extracted and converted to the carrier localization length, $\xi=\frac{2.8 e^{2}}{4 \pi \varepsilon \varepsilon_{0} k_{\mathrm{B}} T_{0}}$~\cite{Efros_1979}, assuming a thickness-independent dielectric constant $\varepsilon = 6.45$~\cite{Bao_2022}. Here $e$ is the elementary charge and $\varepsilon_{0}$ the vacuum permittivity. As shown in Fig.~\ref{Fig3}(f), $\xi$ decreases progressively with thickness reduction, followed by a sharp collapse in bilayer and monolayer samples. The extremely small value for the monolayer (0.36 nm), comparable to the lattice constant~\cite{Lee_2019}, indicates strong carrier localization driven by enhanced electron-electron interactions and increased disorder.

\begin{figure}[t]
\includegraphics[width=0.75\linewidth]{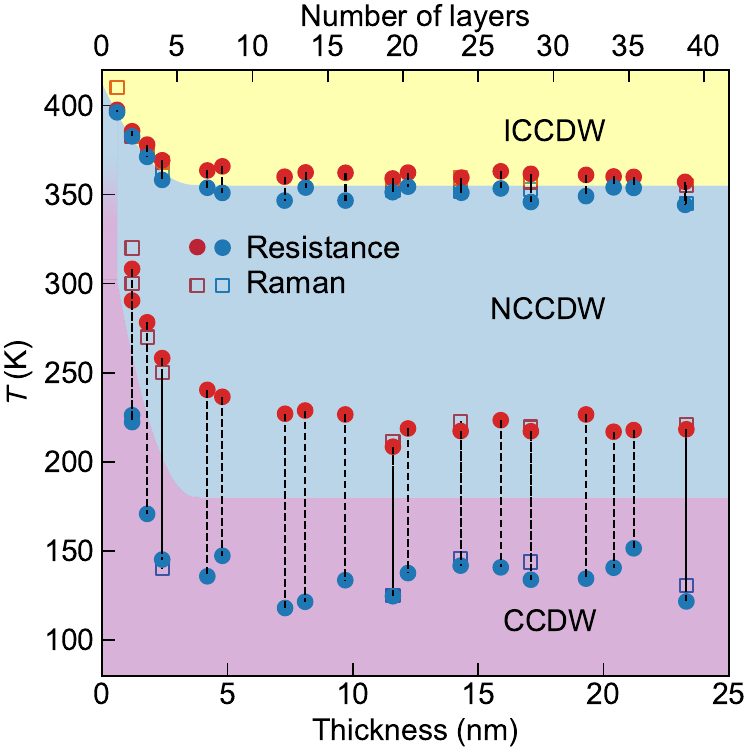} 
\caption{Temperature-thickness phase diagram from transport and Raman measurements. Critical temperatures for the NCCDW-CCDW and ICCDW-NCCDW transitions during cooling and warming are connected by dashed lines.}
\label{Fig4}
\end{figure}

\begin{figure}[t]
\includegraphics[width=\linewidth]{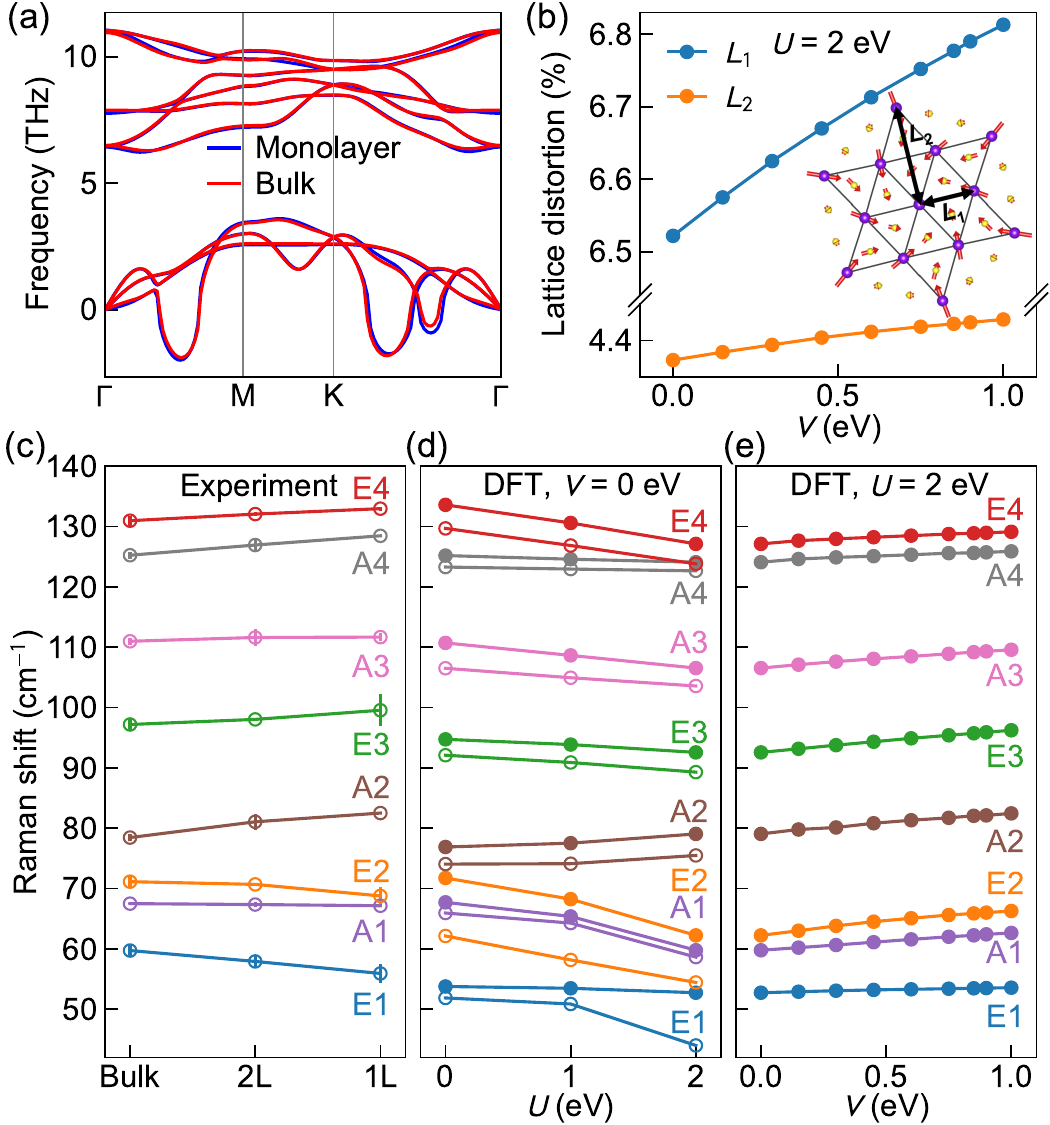} 
\caption{(a) Phonon dispersions of monolayer and bulk 1$T$-TaS$_2$ in the pristine phase ($U=V=0$). (b) Lattice distortion versus $V$ at fixed $U=2$~eV, quantified by the relative changes of the nearest- and next-nearest-neighbor Ta-Ta distances, $L_1$ and $L_2$, with respect to the average bond lengths. The distortion parameters are defined as $(a-L_1)/a$ and $(a-L_2/\sqrt{3})/a$, where $a$ is the average Ta-Ta bond length given by the supercell lattice constant divided by $\sqrt{13}$. Inset: A2 amplitude mode displacement pattern with $L_1$ and $L_2$ marked. (c) Measured phonon frequencies of the A1--A4 and E1--E4 modes for 1L, 2L, and bulk samples at 10~K. The error bars are standard deviations estimated from multiple samples. (d),(e) Corresponding calculated mode frequencies versus $U$ at $V=0$ (d) and versus $V$ at fixed $U=2$~eV (e) for bulk (filled symbols) and monolayer (open symbols) 1$T$-TaS$_2$.}
\label{Fig5}
\end{figure}

We summarize the temperature-thickness phase diagram in Fig.~\ref{Fig4}, combining transition temperatures extracted from Raman and transport measurements (see methods in Supplemental Material, Note~5). Dashed lines connect the critical temperatures of hysteretic first-order transitions. For samples thicker than five layers ($\sim$3 nm), the critical temperatures $T_{\mathrm{CCDW-NCCDW}}$ and $T_{\mathrm{NCCDW-ICCDW}}$, averaged over warming and cooling cycles, are about 180~K and 355~K, respectively. Below five layers, both transition temperatures increase monotonically with decreasing thickness. In the monolayer, $T_{\mathrm{NCCDW-ICCDW}}$ is enhanced by about 50~K compared to the bulk, while $T_{\mathrm{CCDW-NCCDW}}$ cannot be identified due to the absence of clear signatures in Raman or resistance data. Extrapolating the trend observed in 2--4 layer samples suggests that $T_{\mathrm{CCDW-NCCDW}}$ in the monolayer should exceed 300~K, which is consistent with the reported closure of the CCDW gap near 350~K in MBE-grown monolayer $1T$-TaS$_2$~\cite{Chen_2025}. 

To elucidate the dimensionality effect on the CDW, we performed DFT calculations that incorporate the on-site Coulomb interaction $U$ and the in-plane nearest-neighbor inter-site interaction $V$. This is motivated by the expectation that reduced out-of-plane screening in the 2D limit enhances both local and nonlocal Coulomb interactions~\cite{Zhang2024}. Constrained random-phase-approximation calculations yield $U=1.85$~eV and $V=0.55$~eV for the bulk, and $U=2.24$~eV and $V=0.93$~eV for the monolayer, confirming enhanced Coulomb interactions in the 2D limit. Even with $U=V=0$, the phonon dispersion shows a pronounced imaginary branch along $\Gamma$-$M$ at the CDW wavevector [Fig.~\ref{Fig5}(a)], highlighting the key role of electron-phonon coupling in driving the CDW instability. However, the bulk and monolayer dispersions are nearly identical, indicating that standard DFT does not capture the thickness dependence of the CDW. Introducing $U$ softens all A1--A4 and E1--E4 modes except A2 [Fig.~\ref{Fig5}(d)], so an enhanced $U$ alone cannot explain the experimental trends in Fig.~\ref{Fig5}(c), which show softening only for the E1 and E2 modes as the thickness is reduced. At fixed $U$, the monolayer exhibits an overall phonon softening relative to the bulk [Fig.~\ref{Fig5}(d)], indicating reduced lattice rigidity due to the absence of interlayer coupling, consistent with the reduced calculated $c$-axis stiffness of the monolayer (Supplemental Material, Table I).
  
In contrast, at fixed $U=2$~eV, increasing $V$ hardens all modes [Fig.~\ref{Fig5}(e)], demonstrating that $U$, $V$, and lattice rigidity have distinct effects on the lattice dynamics. Their combined influence on the thickness dependence of the phonon frequencies is therefore complex, making quantitative agreement between experiment and calculation difficult. Nevertheless, several key trends emerge from the analysis. First, the experimentally observed hardening of the A3, A4, E3, and E4 modes with decreasing thickness points to a dominant role of enhanced $V$, since both increasing $U$ and reducing lattice rigidity would instead soften these modes. Second, both $U$ and $V$ strengthen the CDW, as reflected by the hardening of the A2 mode, the amplitude mode of the CCDW phase (Supplemental Material, Fig.~8)~\cite{Hasaien2025}, whose breathing-like eigenvector directly modulates the SD cluster [inset in Fig.~\ref{Fig5}(b)]. Consistently, the SD lattice distortion increases with $V$ [Fig.~\ref{Fig5}(b)]. Third, the softening of the E1 and E2 modes in the monolayer is dominated by reduced lattice rigidity. Eigenvector analysis indicates substantial mode mixing between the E1 and E2 modes in the monolayer (Supplemental Material, Fig.~19), consistent with the systematic evolution of their relative Raman intensities (Supplemental Material, Fig.~9), which is not observed for other modes. This mixing redistributes the in-plane and out-of-plane eigenvector components of E1 and E2 and makes them particularly sensitive to interlayer coupling.

Overall, these results support enhanced CDW order in the 2D limit driven by strengthened Coulomb interactions as screening from adjacent layers is reduced, with the nonlocal term $V$ playing the more prominent role. A similar importance of $V$ was reported in 1$T$-NbS$_2$~\cite{Zhang2024}, where it selects the SD distortion among multiple competing lattice instabilities, also evident in Fig.~\ref{Fig5}(a). Likewise, interlayer hopping was shown to reduce correlations and weaken the insulating state in bilayer and trilayer 1$T$-TaSe$_2$, as reflected by a reduced gap and quenched orbital texture~\cite{Chen_2020}.

Both the CDW transition temperatures and the thermal activation gap vary monotonically with thickness in atomically thin samples, which is difficult to reconcile with an interlayer-dimerization driven insulating ground state. For bilayer 1$T$-TaS$_2$, calculations predict AA stacking to be the most stable, with vertically aligned layers forming a dimer~\cite{Jiang_2021,Bae2025}. In this picture, the bilayer hybridization gap is predicted to exceed the Mott gap for a given $U$, in contrast to our observation that the monolayer exhibits both a larger activation gap and higher CDW transition temperatures. For bulk crystals, two stacking configurations with comparable formation energies have been proposed, corresponding to staggered dimers and staggered monolayers along the $c$ axis, respectively; the latter is predicted to be metallic~\cite{Ritschel_2015,Lee_2019}. Such a strong stacking dependence would be expected to produce pronounced sample-to-sample variations in transition temperatures and transport in few-layer samples. In our measurements, however, the spread in transition temperatures at fixed thickness is small compared with the overall monotonic thickness dependence (Supplemental Material, Fig.~16). Moreover, if a metallic layer were present in a few-layer crystal, it should dominate in-plane transport, yet all measured samples are insulating. The observed monotonic increase in CDW transition temperatures and activation gap in atomically thin samples is therefore more consistent with a systematic reduction of screening, which enhances both $U$ and $V$. Although our results are inconsistent with an electronic gap driven by interlayer dimerization, the evidence for structural dimerization remains clear~\cite{Butler2020}. The high-quality atomically thin 1$T$-TaS$_2$ samples reported here provide a platform for investigating how interlayer coupling affects the proposed quantum spin liquid state~\cite{Law2017,Chen_2025,ManasValero2021}.

In conclusion, we combined Raman scattering and electrical transport to probe the thickness dependence of CDW transitions in 1$T$-TaS$_2$. We find that the CCDW phase persists to the monolayer limit, whereas the hysteretic CCDW-NCCDW transition survives down to the bilayer, revealing a clear role of interlayer coupling. Both CCDW-NCCDW and NCCDW-ICCDW transition temperatures increase monotonically with decreasing thickness, consistent with enhanced electron correlations driven by strengthened Coulomb interactions. These results set the stage for tuning correlated insulating phases in atomically thin 1$T$-TaS$_2$ via field-effect gating and heterostructure engineering.

We thank Jian Sun for helpful discussions. This work was supported by the Natural Science Foundation of Jiangsu Province (Grants No. BK20231529, No. BK20233001, and No. BK20253009), the National Natural Science Foundation of China (Grants No. U24A6002, No. 12474170, No. 123B2059, No. 12504212, No. 12225407, No. 12434005, and No. 12274207), the National Key Research and Development Program of China (Grants No. 2024YFA1409100 and No. 2022YFA1403000), the China Postdoctoral Science Foundation (Grant No. BX20240160 and No. 2025M773370), and the Fundamental Research Funds for the Central Universities (Grants No. 0204-14380260 and No. 0204-14380264). K. W. and T. T. acknowledge support from the JSPS KAKENHI (Grants No. 20H00354 and No. 23H02052) and World Premier International Research Center Initiative (WPI), MEXT, Japan.

\textit{Data availability}---The data that support the findings of this article are openly available~\cite{DAS}.


\begin{thebibliography}{51}%
\makeatletter
\providecommand \@ifxundefined [1]{%
 \@ifx{#1\undefined}
}%
\providecommand \@ifnum [1]{%
 \ifnum #1\expandafter \@firstoftwo
 \else \expandafter \@secondoftwo
 \fi
}%
\providecommand \@ifx [1]{%
 \ifx #1\expandafter \@firstoftwo
 \else \expandafter \@secondoftwo
 \fi
}%
\providecommand \natexlab [1]{#1}%
\providecommand \enquote  [1]{``#1''}%
\providecommand \bibnamefont  [1]{#1}%
\providecommand \bibfnamefont [1]{#1}%
\providecommand \citenamefont [1]{#1}%
\providecommand \href@noop [0]{\@secondoftwo}%
\providecommand \href [0]{\begingroup \@sanitize@url \@href}%
\providecommand \@href[1]{\@@startlink{#1}\@@href}%
\providecommand \@@href[1]{\endgroup#1\@@endlink}%
\providecommand \@sanitize@url [0]{\catcode `\\12\catcode `\$12\catcode
  `\&12\catcode `\#12\catcode `\^12\catcode `\_12\catcode `\%12\relax}%
\providecommand \@@startlink[1]{}%
\providecommand \@@endlink[0]{}%
\providecommand \url  [0]{\begingroup\@sanitize@url \@url }%
\providecommand \@url [1]{\endgroup\@href {#1}{\urlprefix }}%
\providecommand \urlprefix  [0]{URL }%
\providecommand \Eprint [0]{\href }%
\providecommand \doibase [0]{https://doi.org/}%
\providecommand \selectlanguage [0]{\@gobble}%
\providecommand \bibinfo  [0]{\@secondoftwo}%
\providecommand \bibfield  [0]{\@secondoftwo}%
\providecommand \translation [1]{[#1]}%
\providecommand \BibitemOpen [0]{}%
\providecommand \bibitemStop [0]{}%
\providecommand \bibitemNoStop [0]{.\EOS\space}%
\providecommand \EOS [0]{\spacefactor3000\relax}%
\providecommand \BibitemShut  [1]{\csname bibitem#1\endcsname}%
\let\auto@bib@innerbib\@empty
\bibitem [{\citenamefont {Gr{\"u}ner}(2018)}]{Gruner2000}%
  \BibitemOpen
  \bibfield  {author} {\bibinfo {author} {\bibfnamefont {G.}~\bibnamefont
  {Gr{\"u}ner}},\ }\href@noop {} {\emph {\bibinfo {title} {{Density Waves in
  Solids}}}}\ (\bibinfo  {publisher} {CRC Press, Boca Raton, Florida},\
  \bibinfo {year} {2018})\BibitemShut {NoStop}%
\bibitem [{\citenamefont {{J. A. Wilson, F. J. Di Salvo, and S.
  Mahajan}}(1975)}]{Wilson_1975}%
  \BibitemOpen
  \bibfield  {author} {\bibinfo {author} {\bibnamefont {{J. A. Wilson, F. J. Di
  Salvo, and S. Mahajan}}},\ }\bibfield  {title} {\bibinfo {title}
  {Charge-density waves and superlattices in the metallic layered transition
  metal dichalcogenides},\ }\href {https://doi.org/10.1080/00018737500101391}
  {\bibfield  {journal} {\bibinfo  {journal} {Adv. Phys.}\ }\textbf {\bibinfo
  {volume} {24}},\ \bibinfo {pages} {117} (\bibinfo {year} {1975})}\BibitemShut
  {NoStop}%
\bibitem [{\citenamefont {Sipos}\ \emph {et~al.}(2008)\citenamefont {Sipos},
  \citenamefont {Kusmartseva}, \citenamefont {Akrap}, \citenamefont {Berger},
  \citenamefont {Forr\'{o}},\ and\ \citenamefont {Tuti\v{s}}}]{Sipos_2008}%
  \BibitemOpen
  \bibfield  {author} {\bibinfo {author} {\bibfnamefont {B.}~\bibnamefont
  {Sipos}}, \bibinfo {author} {\bibfnamefont {A.~F.}\ \bibnamefont
  {Kusmartseva}}, \bibinfo {author} {\bibfnamefont {A.}~\bibnamefont {Akrap}},
  \bibinfo {author} {\bibfnamefont {H.}~\bibnamefont {Berger}}, \bibinfo
  {author} {\bibfnamefont {L.}~\bibnamefont {Forr\'{o}}},\ and\ \bibinfo
  {author} {\bibfnamefont {E.}~\bibnamefont {Tuti\v{s}}},\ }\bibfield  {title}
  {\bibinfo {title} {{From Mott state to superconductivity in 1$T$-TaS$_2$}},\
  }\href {https://doi.org/10.1038/nmat2318} {\bibfield  {journal} {\bibinfo
  {journal} {Nat. Mater.}\ }\textbf {\bibinfo {volume} {7}},\ \bibinfo {pages}
  {960} (\bibinfo {year} {2008})}\BibitemShut {NoStop}%
\bibitem [{\citenamefont {Ang}\ \emph {et~al.}(2012)\citenamefont {Ang},
  \citenamefont {Tanaka}, \citenamefont {Ieki}, \citenamefont {Nakayama},
  \citenamefont {Sato}, \citenamefont {Li}, \citenamefont {Lu}, \citenamefont
  {Sun},\ and\ \citenamefont {Takahashi}}]{Ang_2012}%
  \BibitemOpen
  \bibfield  {author} {\bibinfo {author} {\bibfnamefont {R.}~\bibnamefont
  {Ang}}, \bibinfo {author} {\bibfnamefont {Y.}~\bibnamefont {Tanaka}},
  \bibinfo {author} {\bibfnamefont {E.}~\bibnamefont {Ieki}}, \bibinfo {author}
  {\bibfnamefont {K.}~\bibnamefont {Nakayama}}, \bibinfo {author}
  {\bibfnamefont {T.}~\bibnamefont {Sato}}, \bibinfo {author} {\bibfnamefont
  {L.~J.}\ \bibnamefont {Li}}, \bibinfo {author} {\bibfnamefont {W.~J.}\
  \bibnamefont {Lu}}, \bibinfo {author} {\bibfnamefont {Y.~P.}\ \bibnamefont
  {Sun}},\ and\ \bibinfo {author} {\bibfnamefont {T.}~\bibnamefont
  {Takahashi}},\ }\bibfield  {title} {\bibinfo {title} {{Real-space coexistence
  of the melted Mott state and superconductivity in Fe-substituted
  1$T$-TaS$_2$}},\ }\href {https://doi.org/10.1103/PhysRevLett.109.176403}
  {\bibfield  {journal} {\bibinfo  {journal} {Phys. Rev. Lett.}\ }\textbf
  {\bibinfo {volume} {109}},\ \bibinfo {pages} {176403} (\bibinfo {year}
  {2012})}\BibitemShut {NoStop}%
\bibitem [{\citenamefont {Yu}\ \emph {et~al.}(2015)\citenamefont {Yu},
  \citenamefont {Yang}, \citenamefont {Lu}, \citenamefont {Yan}, \citenamefont
  {Cho}, \citenamefont {Ma}, \citenamefont {Niu}, \citenamefont {Kim},
  \citenamefont {Son}, \citenamefont {Feng}, \citenamefont {Li}, \citenamefont
  {Cheong}, \citenamefont {Chen},\ and\ \citenamefont {Zhang}}]{Yu2015}%
  \BibitemOpen
  \bibfield  {author} {\bibinfo {author} {\bibfnamefont {Y.}~\bibnamefont
  {Yu}}, \bibinfo {author} {\bibfnamefont {F.}~\bibnamefont {Yang}}, \bibinfo
  {author} {\bibfnamefont {X.~F.}\ \bibnamefont {Lu}}, \bibinfo {author}
  {\bibfnamefont {Y.~J.}\ \bibnamefont {Yan}}, \bibinfo {author} {\bibfnamefont
  {Y.-H.}\ \bibnamefont {Cho}}, \bibinfo {author} {\bibfnamefont
  {L.}~\bibnamefont {Ma}}, \bibinfo {author} {\bibfnamefont {X.}~\bibnamefont
  {Niu}}, \bibinfo {author} {\bibfnamefont {S.}~\bibnamefont {Kim}}, \bibinfo
  {author} {\bibfnamefont {Y.-W.}\ \bibnamefont {Son}}, \bibinfo {author}
  {\bibfnamefont {D.}~\bibnamefont {Feng}}, \bibinfo {author} {\bibfnamefont
  {S.}~\bibnamefont {Li}}, \bibinfo {author} {\bibfnamefont {S.-W.}\
  \bibnamefont {Cheong}}, \bibinfo {author} {\bibfnamefont {X.~H.}\
  \bibnamefont {Chen}},\ and\ \bibinfo {author} {\bibfnamefont
  {Y.}~\bibnamefont {Zhang}},\ }\bibfield  {title} {\bibinfo {title}
  {{Gate-tunable phase transitions in thin flakes of 1$T$-TaS$_2$}},\ }\href
  {https://doi.org/10.1038/nnano.2014.323} {\bibfield  {journal} {\bibinfo
  {journal} {Nat. Nanotechnol.}\ }\textbf {\bibinfo {volume} {10}},\ \bibinfo
  {pages} {270} (\bibinfo {year} {2015})}\BibitemShut {NoStop}%
\bibitem [{\citenamefont {Dong}\ \emph {et~al.}(2021)\citenamefont {Dong},
  \citenamefont {Li}, \citenamefont {Li}, \citenamefont {Shi}, \citenamefont
  {Niu}, \citenamefont {Liu}, \citenamefont {Liu}, \citenamefont {Liu},
  \citenamefont {Luo}, \citenamefont {Si}, \citenamefont {Lu}, \citenamefont
  {Hao}, \citenamefont {Sun},\ and\ \citenamefont {Liu}}]{Dong_2021}%
  \BibitemOpen
  \bibfield  {author} {\bibinfo {author} {\bibfnamefont {Q.}~\bibnamefont
  {Dong}}, \bibinfo {author} {\bibfnamefont {Q.}~\bibnamefont {Li}}, \bibinfo
  {author} {\bibfnamefont {S.}~\bibnamefont {Li}}, \bibinfo {author}
  {\bibfnamefont {X.}~\bibnamefont {Shi}}, \bibinfo {author} {\bibfnamefont
  {S.}~\bibnamefont {Niu}}, \bibinfo {author} {\bibfnamefont {S.}~\bibnamefont
  {Liu}}, \bibinfo {author} {\bibfnamefont {R.}~\bibnamefont {Liu}}, \bibinfo
  {author} {\bibfnamefont {B.}~\bibnamefont {Liu}}, \bibinfo {author}
  {\bibfnamefont {X.}~\bibnamefont {Luo}}, \bibinfo {author} {\bibfnamefont
  {J.}~\bibnamefont {Si}}, \bibinfo {author} {\bibfnamefont {W.}~\bibnamefont
  {Lu}}, \bibinfo {author} {\bibfnamefont {N.}~\bibnamefont {Hao}}, \bibinfo
  {author} {\bibfnamefont {Y.}~\bibnamefont {Sun}},\ and\ \bibinfo {author}
  {\bibfnamefont {B.}~\bibnamefont {Liu}},\ }\bibfield  {title} {\bibinfo
  {title} {{Structural phase transition and superconductivity hierarchy in
  1$T$-TaS$_2$ under pressure up to 100 GPa}},\ }\href
  {https://doi.org/10.1038/s41535-021-00320-x} {\bibfield  {journal} {\bibinfo
  {journal} {npj Quantum Mater}\ }\textbf {\bibinfo {volume} {6}},\ \bibinfo
  {pages} {20} (\bibinfo {year} {2021})}\BibitemShut {NoStop}%
\bibitem [{\citenamefont {Law}\ and\ \citenamefont {Lee}(2017)}]{Law2017}%
  \BibitemOpen
  \bibfield  {author} {\bibinfo {author} {\bibfnamefont {K.~T.}\ \bibnamefont
  {Law}}\ and\ \bibinfo {author} {\bibfnamefont {P.~A.}\ \bibnamefont {Lee}},\
  }\bibfield  {title} {\bibinfo {title} {{1$T$-TaS$_2$ as a quantum spin
  liquid}},\ }\href {https://doi.org/10.1073/pnas.1706769114} {\bibfield
  {journal} {\bibinfo  {journal} {Proc. Natl. Acad. Sci. U.S.A.}\ }\textbf
  {\bibinfo {volume} {114}},\ \bibinfo {pages} {6996} (\bibinfo {year}
  {2017})}\BibitemShut {NoStop}%
\bibitem [{\citenamefont {Klanj\v{s}ek}\ \emph {et~al.}(2017)\citenamefont
  {Klanj\v{s}ek}, \citenamefont {Zorko}, \citenamefont {\v{Z}itko},
  \citenamefont {Mravlje}, \citenamefont {Jagli\v{c}i\'{c}}, \citenamefont {{P.
  K. Biswas}}, \citenamefont {Prelov\v{s}ek}, \citenamefont {Mihailovic},\ and\
  \citenamefont {Ar\v{c}on}}]{Martin_2017}%
  \BibitemOpen
  \bibfield  {author} {\bibinfo {author} {\bibfnamefont {M.}~\bibnamefont
  {Klanj\v{s}ek}}, \bibinfo {author} {\bibfnamefont {A.}~\bibnamefont {Zorko}},
  \bibinfo {author} {\bibfnamefont {R.}~\bibnamefont {\v{Z}itko}}, \bibinfo
  {author} {\bibfnamefont {J.}~\bibnamefont {Mravlje}}, \bibinfo {author}
  {\bibfnamefont {Z.}~\bibnamefont {Jagli\v{c}i\'{c}}}, \bibinfo {author}
  {\bibnamefont {{P. K. Biswas}}}, \bibinfo {author} {\bibfnamefont
  {P.}~\bibnamefont {Prelov\v{s}ek}}, \bibinfo {author} {\bibfnamefont
  {D.}~\bibnamefont {Mihailovic}},\ and\ \bibinfo {author} {\bibfnamefont
  {D.}~\bibnamefont {Ar\v{c}on}},\ }\bibfield  {title} {\bibinfo {title} {{A
  high-temperature quantum spin liquid with polaron spins}},\ }\href
  {https://doi.org/10.1038/nphys4212} {\bibfield  {journal} {\bibinfo
  {journal} {Nat. Phys.}\ }\textbf {\bibinfo {volume} {13}},\ \bibinfo {pages}
  {1130} (\bibinfo {year} {2017})}\BibitemShut {NoStop}%
\bibitem [{\citenamefont {He}\ \emph {et~al.}(2018)\citenamefont {He},
  \citenamefont {Xu}, \citenamefont {Chen}, \citenamefont {Law},\ and\
  \citenamefont {Lee}}]{He_2018}%
  \BibitemOpen
  \bibfield  {author} {\bibinfo {author} {\bibfnamefont {W.-Y.}\ \bibnamefont
  {He}}, \bibinfo {author} {\bibfnamefont {X.~Y.}\ \bibnamefont {Xu}}, \bibinfo
  {author} {\bibfnamefont {G.}~\bibnamefont {Chen}}, \bibinfo {author}
  {\bibfnamefont {K.~T.}\ \bibnamefont {Law}},\ and\ \bibinfo {author}
  {\bibfnamefont {P.~A.}\ \bibnamefont {Lee}},\ }\bibfield  {title} {\bibinfo
  {title} {{Spinon Fermi surface in a cluster Mott insulator model on a
  triangular lattice and possible application to 1$T$-TaS$_2$}},\ }\href
  {https://doi.org/10.1103/PhysRevLett.121.046401} {\bibfield  {journal}
  {\bibinfo  {journal} {Phys. Rev. Lett.}\ }\textbf {\bibinfo {volume} {121}},\
  \bibinfo {pages} {046401} (\bibinfo {year} {2018})}\BibitemShut {NoStop}%
\bibitem [{\citenamefont {Woolley}\ and\ \citenamefont
  {Wexler}(1977)}]{Woolley_1977}%
  \BibitemOpen
  \bibfield  {author} {\bibinfo {author} {\bibfnamefont {A.~M.}\ \bibnamefont
  {Woolley}}\ and\ \bibinfo {author} {\bibfnamefont {G.}~\bibnamefont
  {Wexler}},\ }\bibfield  {title} {\bibinfo {title} {{Band structures and Fermi
  surfaces for 1$T$-TaS$_2$, 1$T$-TaSe$_2$ and 1$T$-VSe$_2$}},\ }\href
  {https://doi.org/10.1088/0022-3719/10/14/013} {\bibfield  {journal} {\bibinfo
   {journal} {J. Phys. C}\ }\textbf {\bibinfo {volume} {10}},\ \bibinfo {pages}
  {2601} (\bibinfo {year} {1977})}\BibitemShut {NoStop}%
\bibitem [{\citenamefont {Myron}\ and\ \citenamefont
  {Freeman}(1975)}]{PMyron_1975}%
  \BibitemOpen
  \bibfield  {author} {\bibinfo {author} {\bibfnamefont {H.~W.}\ \bibnamefont
  {Myron}}\ and\ \bibinfo {author} {\bibfnamefont {A.~J.}\ \bibnamefont
  {Freeman}},\ }\bibfield  {title} {\bibinfo {title} {{Electronic structure and
  Fermi-surface-related instabilities in 1$T$-TaS$_2$ and 1$T$-TaSe$_2$}},\
  }\href {https://doi.org/10.1103/PhysRevB.11.2735} {\bibfield  {journal}
  {\bibinfo  {journal} {Phys. Rev. B}\ }\textbf {\bibinfo {volume} {11}},\
  \bibinfo {pages} {2735} (\bibinfo {year} {1975})}\BibitemShut {NoStop}%
\bibitem [{\citenamefont {Bovet}\ \emph {et~al.}(2004)\citenamefont {Bovet},
  \citenamefont {Popovi\ifmmode~\acute{c}\else \'{c}\fi{}}, \citenamefont
  {Clerc}, \citenamefont {Koitzsch}, \citenamefont {Probst}, \citenamefont
  {Bucher}, \citenamefont {Berger}, \citenamefont
  {Naumovi\ifmmode~\acute{c}\else \'{c}\fi{}},\ and\ \citenamefont
  {Aebi}}]{Bovet_2004}%
  \BibitemOpen
  \bibfield  {author} {\bibinfo {author} {\bibfnamefont {M.}~\bibnamefont
  {Bovet}}, \bibinfo {author} {\bibfnamefont {D.}~\bibnamefont
  {Popovi\ifmmode~\acute{c}\else \'{c}\fi{}}}, \bibinfo {author} {\bibfnamefont
  {F.}~\bibnamefont {Clerc}}, \bibinfo {author} {\bibfnamefont
  {C.}~\bibnamefont {Koitzsch}}, \bibinfo {author} {\bibfnamefont
  {U.}~\bibnamefont {Probst}}, \bibinfo {author} {\bibfnamefont
  {E.}~\bibnamefont {Bucher}}, \bibinfo {author} {\bibfnamefont
  {H.}~\bibnamefont {Berger}}, \bibinfo {author} {\bibfnamefont
  {D.}~\bibnamefont {Naumovi\ifmmode~\acute{c}\else \'{c}\fi{}}},\ and\
  \bibinfo {author} {\bibfnamefont {P.}~\bibnamefont {Aebi}},\ }\bibfield
  {title} {\bibinfo {title} {{Pseudogapped Fermi surfaces of 1$T$-TaS$_2$ and
  1$T$-TaSe$_2$: A charge density wave effect}},\ }\href
  {https://doi.org/10.1103/PhysRevB.69.125117} {\bibfield  {journal} {\bibinfo
  {journal} {Phys. Rev. B}\ }\textbf {\bibinfo {volume} {69}},\ \bibinfo
  {pages} {125117} (\bibinfo {year} {2004})}\BibitemShut {NoStop}%
\bibitem [{\citenamefont {Clerc}\ \emph {et~al.}(2006)\citenamefont {Clerc},
  \citenamefont {Battaglia}, \citenamefont {Bovet}, \citenamefont {Despont},
  \citenamefont {Monney}, \citenamefont {Cercellier}, \citenamefont {Garnier},
  \citenamefont {Aebi}, \citenamefont {Berger},\ and\ \citenamefont
  {Forr\'o}}]{Clerc_2006}%
  \BibitemOpen
  \bibfield  {author} {\bibinfo {author} {\bibfnamefont {F.}~\bibnamefont
  {Clerc}}, \bibinfo {author} {\bibfnamefont {C.}~\bibnamefont {Battaglia}},
  \bibinfo {author} {\bibfnamefont {M.}~\bibnamefont {Bovet}}, \bibinfo
  {author} {\bibfnamefont {L.}~\bibnamefont {Despont}}, \bibinfo {author}
  {\bibfnamefont {C.}~\bibnamefont {Monney}}, \bibinfo {author} {\bibfnamefont
  {H.}~\bibnamefont {Cercellier}}, \bibinfo {author} {\bibfnamefont {M.~G.}\
  \bibnamefont {Garnier}}, \bibinfo {author} {\bibfnamefont {P.}~\bibnamefont
  {Aebi}}, \bibinfo {author} {\bibfnamefont {H.}~\bibnamefont {Berger}},\ and\
  \bibinfo {author} {\bibfnamefont {L.}~\bibnamefont {Forr\'o}},\ }\bibfield
  {title} {\bibinfo {title} {{Lattice-distortion-enhanced electron-phonon
  coupling and Fermi surface nesting in 1$T$-TaS$_2$}},\ }\href
  {https://doi.org/10.1103/PhysRevB.74.155114} {\bibfield  {journal} {\bibinfo
  {journal} {Phys. Rev. B}\ }\textbf {\bibinfo {volume} {74}},\ \bibinfo
  {pages} {155114} (\bibinfo {year} {2006})}\BibitemShut {NoStop}%
\bibitem [{\citenamefont {Liu}(2009)}]{Liu_2009}%
  \BibitemOpen
  \bibfield  {author} {\bibinfo {author} {\bibfnamefont {A.~Y.}\ \bibnamefont
  {Liu}},\ }\bibfield  {title} {\bibinfo {title} {{Electron-phonon coupling in
  compressed 1$T$-TaS$_2$: Stability and superconductivity from first
  principles}},\ }\href {https://doi.org/10.1103/PhysRevB.79.220515} {\bibfield
   {journal} {\bibinfo  {journal} {Phys. Rev. B}\ }\textbf {\bibinfo {volume}
  {79}},\ \bibinfo {pages} {220515 (R)} (\bibinfo {year} {2009})}\BibitemShut
  {NoStop}%
\bibitem [{\citenamefont {Rossnagel}(2011)}]{Rossnagel_2011}%
  \BibitemOpen
  \bibfield  {author} {\bibinfo {author} {\bibfnamefont {K.}~\bibnamefont
  {Rossnagel}},\ }\bibfield  {title} {\bibinfo {title} {{On the origin of
  charge-density waves in select layered transition-metal dichalcogenides}},\
  }\href {https://doi.org/10.1088/0953-8984/23/21/213001} {\bibfield  {journal}
  {\bibinfo  {journal} {J. Condens.: Matter Phys.}\ }\textbf {\bibinfo {volume}
  {23}},\ \bibinfo {pages} {213001} (\bibinfo {year} {2011})}\BibitemShut
  {NoStop}%
\bibitem [{\citenamefont {Zhang}\ \emph {et~al.}(2014)\citenamefont {Zhang},
  \citenamefont {Gan}, \citenamefont {Cheng},\ and\ \citenamefont
  {Schwingenschl\"ogl}}]{Zhang_2014}%
  \BibitemOpen
  \bibfield  {author} {\bibinfo {author} {\bibfnamefont {Q.}~\bibnamefont
  {Zhang}}, \bibinfo {author} {\bibfnamefont {L.-Y.}\ \bibnamefont {Gan}},
  \bibinfo {author} {\bibfnamefont {Y.}~\bibnamefont {Cheng}},\ and\ \bibinfo
  {author} {\bibfnamefont {U.}~\bibnamefont {Schwingenschl\"ogl}},\ }\bibfield
  {title} {\bibinfo {title} {{Spin polarization driven by a charge-density wave
  in monolayer 1$T$-TaS$_2$}},\ }\href
  {https://doi.org/10.1103/PhysRevB.90.081103} {\bibfield  {journal} {\bibinfo
  {journal} {Phys. Rev. B}\ }\textbf {\bibinfo {volume} {90}},\ \bibinfo
  {pages} {081103 (R)} (\bibinfo {year} {2014})}\BibitemShut {NoStop}%
\bibitem [{\citenamefont {Cho}\ \emph {et~al.}(2015)\citenamefont {Cho},
  \citenamefont {Cho}, \citenamefont {Cheong}, \citenamefont {Kim},\ and\
  \citenamefont {Yeom}}]{Cho_2015}%
  \BibitemOpen
  \bibfield  {author} {\bibinfo {author} {\bibfnamefont {D.}~\bibnamefont
  {Cho}}, \bibinfo {author} {\bibfnamefont {Y.-H.}\ \bibnamefont {Cho}},
  \bibinfo {author} {\bibfnamefont {S.-W.}\ \bibnamefont {Cheong}}, \bibinfo
  {author} {\bibfnamefont {K.-S.}\ \bibnamefont {Kim}},\ and\ \bibinfo {author}
  {\bibfnamefont {H.~W.}\ \bibnamefont {Yeom}},\ }\bibfield  {title} {\bibinfo
  {title} {{Interplay of electron-electron and electron-phonon interactions in
  the low-temperature phase of 1$T$-TaS$_2$}},\ }\href
  {https://doi.org/10.1103/PhysRevB.92.085132} {\bibfield  {journal} {\bibinfo
  {journal} {Phys. Rev. B}\ }\textbf {\bibinfo {volume} {92}},\ \bibinfo
  {pages} {085132} (\bibinfo {year} {2015})}\BibitemShut {NoStop}%
\bibitem [{\citenamefont {Yi}\ \emph {et~al.}(2018)\citenamefont {Yi},
  \citenamefont {Zhang},\ and\ \citenamefont {Cho}}]{Yi_2018}%
  \BibitemOpen
  \bibfield  {author} {\bibinfo {author} {\bibfnamefont {S.}~\bibnamefont
  {Yi}}, \bibinfo {author} {\bibfnamefont {Z.}~\bibnamefont {Zhang}},\ and\
  \bibinfo {author} {\bibfnamefont {J.-H.}\ \bibnamefont {Cho}},\ }\bibfield
  {title} {\bibinfo {title} {{Coupling of charge, lattice, orbital, and spin
  degrees of freedom in charge density waves in 1$T$-TaS$_2$}},\ }\href
  {https://doi.org/10.1103/PhysRevB.97.041413} {\bibfield  {journal} {\bibinfo
  {journal} {Phys. Rev. B}\ }\textbf {\bibinfo {volume} {97}},\ \bibinfo
  {pages} {041413 (R)} (\bibinfo {year} {2018})}\BibitemShut {NoStop}%
\bibitem [{\citenamefont {Fazekas}\ and\ \citenamefont
  {Tosatti}(1980)}]{Fazekas_1980}%
  \BibitemOpen
  \bibfield  {author} {\bibinfo {author} {\bibfnamefont {P.}~\bibnamefont
  {Fazekas}}\ and\ \bibinfo {author} {\bibfnamefont {E.}~\bibnamefont
  {Tosatti}},\ }\bibfield  {title} {\bibinfo {title} {{Charge carrier
  localization in pure and doped 1$T$-TaS$_2$}},\ }\href
  {https://doi.org/10.1016/0378-4363(80)90229-6} {\bibfield  {journal}
  {\bibinfo  {journal} {Physica B+C}\ }\textbf {\bibinfo {volume} {99}},\
  \bibinfo {pages} {183} (\bibinfo {year} {1980})}\BibitemShut {NoStop}%
\bibitem [{\citenamefont {Ritschel}\ \emph {et~al.}(2015)\citenamefont
  {Ritschel}, \citenamefont {Trinckauf}, \citenamefont {Koepernik},
  \citenamefont {Büchner}, \citenamefont {Zimmermann}, \citenamefont {Berger},
  \citenamefont {Joe}, \citenamefont {Abbamonte},\ and\ \citenamefont
  {Geck}}]{Ritschel_2015}%
  \BibitemOpen
  \bibfield  {author} {\bibinfo {author} {\bibfnamefont {T.}~\bibnamefont
  {Ritschel}}, \bibinfo {author} {\bibfnamefont {J.}~\bibnamefont {Trinckauf}},
  \bibinfo {author} {\bibfnamefont {K.}~\bibnamefont {Koepernik}}, \bibinfo
  {author} {\bibfnamefont {B.}~\bibnamefont {Büchner}}, \bibinfo {author}
  {\bibfnamefont {M.~v.}\ \bibnamefont {Zimmermann}}, \bibinfo {author}
  {\bibfnamefont {H.}~\bibnamefont {Berger}}, \bibinfo {author} {\bibfnamefont
  {Y.~I.}\ \bibnamefont {Joe}}, \bibinfo {author} {\bibfnamefont
  {P.}~\bibnamefont {Abbamonte}},\ and\ \bibinfo {author} {\bibfnamefont
  {J.}~\bibnamefont {Geck}},\ }\bibfield  {title} {\bibinfo {title} {{Orbital
  textures and charge density waves in transition metal dichalcogenides}},\
  }\href {https://doi.org/10.1038/nphys3267} {\bibfield  {journal} {\bibinfo
  {journal} {Nat. Phys.}\ }\textbf {\bibinfo {volume} {11}},\ \bibinfo {pages}
  {328} (\bibinfo {year} {2015})}\BibitemShut {NoStop}%
\bibitem [{\citenamefont {Lee}\ \emph {et~al.}(2019)\citenamefont {Lee},
  \citenamefont {Goh},\ and\ \citenamefont {Cho}}]{Lee_2019}%
  \BibitemOpen
  \bibfield  {author} {\bibinfo {author} {\bibfnamefont {S.-H.}\ \bibnamefont
  {Lee}}, \bibinfo {author} {\bibfnamefont {J.~S.}\ \bibnamefont {Goh}},\ and\
  \bibinfo {author} {\bibfnamefont {D.}~\bibnamefont {Cho}},\ }\bibfield
  {title} {\bibinfo {title} {{Origin of the insulating phase and first-order
  metal-insulator transition in 1$T$-TaS$_2$}},\ }\href
  {https://doi.org/10.1103/PhysRevLett.122.106404} {\bibfield  {journal}
  {\bibinfo  {journal} {Phys. Rev. Lett.}\ }\textbf {\bibinfo {volume} {122}},\
  \bibinfo {pages} {106404} (\bibinfo {year} {2019})}\BibitemShut {NoStop}%
\bibitem [{\citenamefont {Albertini}\ \emph {et~al.}(2016)\citenamefont
  {Albertini}, \citenamefont {Zhao}, \citenamefont {McCann}, \citenamefont
  {Feng}, \citenamefont {Terrones}, \citenamefont {Freericks}, \citenamefont
  {Robinson},\ and\ \citenamefont {Liu}}]{Albertini_2016}%
  \BibitemOpen
  \bibfield  {author} {\bibinfo {author} {\bibfnamefont {O.~R.}\ \bibnamefont
  {Albertini}}, \bibinfo {author} {\bibfnamefont {R.}~\bibnamefont {Zhao}},
  \bibinfo {author} {\bibfnamefont {R.~L.}\ \bibnamefont {McCann}}, \bibinfo
  {author} {\bibfnamefont {S.}~\bibnamefont {Feng}}, \bibinfo {author}
  {\bibfnamefont {M.}~\bibnamefont {Terrones}}, \bibinfo {author}
  {\bibfnamefont {J.~K.}\ \bibnamefont {Freericks}}, \bibinfo {author}
  {\bibfnamefont {J.~A.}\ \bibnamefont {Robinson}},\ and\ \bibinfo {author}
  {\bibfnamefont {A.~Y.}\ \bibnamefont {Liu}},\ }\bibfield  {title} {\bibinfo
  {title} {{Zone-center phonons of bulk, few-layer, and monolayer 1$T$-TaS$_2$:
  Detection of commensurate charge density wave phase through Raman
  scattering}},\ }\href {https://doi.org/10.1103/PhysRevB.93.214109} {\bibfield
   {journal} {\bibinfo  {journal} {Phys. Rev. B}\ }\textbf {\bibinfo {volume}
  {93}},\ \bibinfo {pages} {214109} (\bibinfo {year} {2016})}\BibitemShut
  {NoStop}%
\bibitem [{\citenamefont {He}\ \emph {et~al.}(2016)\citenamefont {He},
  \citenamefont {Okamoto}, \citenamefont {Ye}, \citenamefont {Ye},
  \citenamefont {Anderson}, \citenamefont {Dai}, \citenamefont {Wu},
  \citenamefont {Hu}, \citenamefont {Liu}, \citenamefont {Lu}, \citenamefont
  {Sun}, \citenamefont {Pasupathy},\ and\ \citenamefont {Tsen}}]{He2016}%
  \BibitemOpen
  \bibfield  {author} {\bibinfo {author} {\bibfnamefont {R.}~\bibnamefont
  {He}}, \bibinfo {author} {\bibfnamefont {J.}~\bibnamefont {Okamoto}},
  \bibinfo {author} {\bibfnamefont {Z.}~\bibnamefont {Ye}}, \bibinfo {author}
  {\bibfnamefont {G.}~\bibnamefont {Ye}}, \bibinfo {author} {\bibfnamefont
  {H.}~\bibnamefont {Anderson}}, \bibinfo {author} {\bibfnamefont
  {X.}~\bibnamefont {Dai}}, \bibinfo {author} {\bibfnamefont {X.}~\bibnamefont
  {Wu}}, \bibinfo {author} {\bibfnamefont {J.}~\bibnamefont {Hu}}, \bibinfo
  {author} {\bibfnamefont {Y.}~\bibnamefont {Liu}}, \bibinfo {author}
  {\bibfnamefont {W.}~\bibnamefont {Lu}}, \bibinfo {author} {\bibfnamefont
  {Y.}~\bibnamefont {Sun}}, \bibinfo {author} {\bibfnamefont {A.~N.}\
  \bibnamefont {Pasupathy}},\ and\ \bibinfo {author} {\bibfnamefont {A.~W.}\
  \bibnamefont {Tsen}},\ }\bibfield  {title} {\bibinfo {title} {Distinct
  surface and bulk charge density waves in ultrathin {1$T$-TaS$_2$}},\ }\href
  {https://doi.org/10.1103/PhysRevB.94.201108} {\bibfield  {journal} {\bibinfo
  {journal} {Phys. Rev. B}\ }\textbf {\bibinfo {volume} {94}},\ \bibinfo
  {pages} {201108 (R)} (\bibinfo {year} {2016})}\BibitemShut {NoStop}%
\bibitem [{\citenamefont {Lin}\ \emph {et~al.}(2020)\citenamefont {Lin},
  \citenamefont {Huang}, \citenamefont {Zhao}, \citenamefont {Qiao},
  \citenamefont {Liu}, \citenamefont {Wu}, \citenamefont {Chen},\ and\
  \citenamefont {Ji}}]{Lin_2020}%
  \BibitemOpen
  \bibfield  {author} {\bibinfo {author} {\bibfnamefont {H.}~\bibnamefont
  {Lin}}, \bibinfo {author} {\bibfnamefont {W.}~\bibnamefont {Huang}}, \bibinfo
  {author} {\bibfnamefont {K.}~\bibnamefont {Zhao}}, \bibinfo {author}
  {\bibfnamefont {S.}~\bibnamefont {Qiao}}, \bibinfo {author} {\bibfnamefont
  {Z.}~\bibnamefont {Liu}}, \bibinfo {author} {\bibfnamefont {J.}~\bibnamefont
  {Wu}}, \bibinfo {author} {\bibfnamefont {X.}~\bibnamefont {Chen}},\ and\
  \bibinfo {author} {\bibfnamefont {S.-H.}\ \bibnamefont {Ji}},\ }\bibfield
  {title} {\bibinfo {title} {{Scanning tunneling spectroscopic study of
  monolayer 1$T$-TaS$_2$ and 1$T$-TaSe$_2$}},\ }\href
  {https://doi.org/10.1007/s12274-019-2584-4} {\bibfield  {journal} {\bibinfo
  {journal} {Nano Res.}\ }\textbf {\bibinfo {volume} {13}},\ \bibinfo {pages}
  {133} (\bibinfo {year} {2020})}\BibitemShut {NoStop}%
\bibitem [{\citenamefont {Chen}\ \emph {et~al.}(2025)\citenamefont {Chen},
  \citenamefont {Wang}, \citenamefont {Gao}, \citenamefont {Gao}, \citenamefont
  {Chen}, \citenamefont {Huang}, \citenamefont {Law}, \citenamefont {Xu},\ and\
  \citenamefont {Chen}}]{Chen_2025}%
  \BibitemOpen
  \bibfield  {author} {\bibinfo {author} {\bibfnamefont {H.}~\bibnamefont
  {Chen}}, \bibinfo {author} {\bibfnamefont {F.-H.}\ \bibnamefont {Wang}},
  \bibinfo {author} {\bibfnamefont {Q.}~\bibnamefont {Gao}}, \bibinfo {author}
  {\bibfnamefont {X.-J.}\ \bibnamefont {Gao}}, \bibinfo {author} {\bibfnamefont
  {Z.}~\bibnamefont {Chen}}, \bibinfo {author} {\bibfnamefont {Y.}~\bibnamefont
  {Huang}}, \bibinfo {author} {\bibfnamefont {K.~T.}\ \bibnamefont {Law}},
  \bibinfo {author} {\bibfnamefont {X.~Y.}\ \bibnamefont {Xu}},\ and\ \bibinfo
  {author} {\bibfnamefont {P.}~\bibnamefont {Chen}},\ }\bibfield  {title}
  {\bibinfo {title} {Spectroscopic evidence for possible quantum spin liquid
  behavior in a two-dimensional {M}ott insulator},\ }\href
  {https://doi.org/10.1103/PhysRevLett.134.066402} {\bibfield  {journal}
  {\bibinfo  {journal} {Phys. Rev. Lett.}\ }\textbf {\bibinfo {volume} {134}},\
  \bibinfo {pages} {066402} (\bibinfo {year} {2025})}\BibitemShut {NoStop}%
\bibitem [{Note1()}]{Note1}%
  \BibitemOpen
  \bibinfo {note} {See Supplemental Material for experimental methods (sample
  preparation and device fabrication, thickness determination, transport and
  optical measurements), Raman mode fitting analysis, additional Raman data,
  analysis of sheet resistance and critical temperatures, and additional
  calculation results.}\BibitemShut {Stop}%
\bibitem [{\citenamefont {Mattheiss}(1973)}]{Mattheiss_1973}%
  \BibitemOpen
  \bibfield  {author} {\bibinfo {author} {\bibfnamefont {L.~F.}\ \bibnamefont
  {Mattheiss}},\ }\bibfield  {title} {\bibinfo {title} {Band structures of
  transition-metal-dichalcogenide layer compounds},\ }\href
  {https://doi.org/10.1103/PhysRevB.8.3719} {\bibfield  {journal} {\bibinfo
  {journal} {Phys. Rev. B}\ }\textbf {\bibinfo {volume} {8}},\ \bibinfo {pages}
  {3719} (\bibinfo {year} {1973})}\BibitemShut {NoStop}%
\bibitem [{\citenamefont {{H. F. Yang $et~al$.}}(2022)}]{Yang2022}%
  \BibitemOpen
  \bibfield  {author} {\bibinfo {author} {\bibnamefont {{H. F. Yang
  $et~al$.}}},\ }\bibfield  {title} {\bibinfo {title} {Visualization of chiral
  electronic structure and anomalous optical response in a material with chiral
  charge density waves},\ }\href
  {https://doi.org/10.1103/PhysRevLett.129.156401} {\bibfield  {journal}
  {\bibinfo  {journal} {Phys. Rev. Lett.}\ }\textbf {\bibinfo {volume} {129}},\
  \bibinfo {pages} {156401} (\bibinfo {year} {2022})}\BibitemShut {NoStop}%
\bibitem [{\citenamefont {Liu}\ \emph {et~al.}(2023)\citenamefont {Liu},
  \citenamefont {Qiu}, \citenamefont {He}, \citenamefont {Liu}, \citenamefont
  {Lin}, \citenamefont {Ma}, \citenamefont {Huang}, \citenamefont {Tang},
  \citenamefont {Xu}, \citenamefont {Watanabe}, \citenamefont {Taniguchi},
  \citenamefont {Gao}, \citenamefont {Wen}, \citenamefont {Liu}, \citenamefont
  {Yan},\ and\ \citenamefont {Xi}}]{Liu_2023}%
  \BibitemOpen
  \bibfield  {author} {\bibinfo {author} {\bibfnamefont {G.}~\bibnamefont
  {Liu}}, \bibinfo {author} {\bibfnamefont {T.}~\bibnamefont {Qiu}}, \bibinfo
  {author} {\bibfnamefont {K.}~\bibnamefont {He}}, \bibinfo {author}
  {\bibfnamefont {Y.}~\bibnamefont {Liu}}, \bibinfo {author} {\bibfnamefont
  {D.}~\bibnamefont {Lin}}, \bibinfo {author} {\bibfnamefont {Z.}~\bibnamefont
  {Ma}}, \bibinfo {author} {\bibfnamefont {Z.}~\bibnamefont {Huang}}, \bibinfo
  {author} {\bibfnamefont {W.}~\bibnamefont {Tang}}, \bibinfo {author}
  {\bibfnamefont {J.}~\bibnamefont {Xu}}, \bibinfo {author} {\bibfnamefont
  {K.}~\bibnamefont {Watanabe}}, \bibinfo {author} {\bibfnamefont
  {T.}~\bibnamefont {Taniguchi}}, \bibinfo {author} {\bibfnamefont
  {L.}~\bibnamefont {Gao}}, \bibinfo {author} {\bibfnamefont {J.}~\bibnamefont
  {Wen}}, \bibinfo {author} {\bibfnamefont {J.-M.}\ \bibnamefont {Liu}},
  \bibinfo {author} {\bibfnamefont {B.}~\bibnamefont {Yan}},\ and\ \bibinfo
  {author} {\bibfnamefont {X.}~\bibnamefont {Xi}},\ }\bibfield  {title}
  {\bibinfo {title} {{Electrical switching of ferro-rotational order in
  nanometre-thick 1$T$-TaS$_2$ crystals}},\ }\href
  {https://doi.org/10.1038/s41565-023-01403-5} {\bibfield  {journal} {\bibinfo
  {journal} {Nat. Nanotechnol.}\ }\textbf {\bibinfo {volume} {18}},\ \bibinfo
  {pages} {854} (\bibinfo {year} {2023})}\BibitemShut {NoStop}%
\bibitem [{\citenamefont {Djurdji\'{c}~Mijin}\ \emph
  {et~al.}(2021)\citenamefont {Djurdji\'{c}~Mijin}, \citenamefont {Baum},
  \citenamefont {Bekaert}, \citenamefont {\ifmmode \check{S}\else
  \v{S}\fi{}olaji\ifmmode~\acute{c}\else \'{c}\fi{}}, \citenamefont {Pe\ifmmode
  \check{s}\else \v{s}\fi{}i\ifmmode~\acute{c}\else \'{c}\fi{}}, \citenamefont
  {Liu}, \citenamefont {He}, \citenamefont {Milo\ifmmode \check{s}\else
  \v{s}\fi{}evi\ifmmode~\acute{c}\else \'{c}\fi{}}, \citenamefont {Petrovic},
  \citenamefont {Popovi\ifmmode~\acute{c}\else \'{c}\fi{}}, \citenamefont
  {Hackl},\ and\ \citenamefont {Lazarevi\ifmmode~\acute{c}\else
  \'{c}\fi{}}}]{Mijin_2021}%
  \BibitemOpen
  \bibfield  {author} {\bibinfo {author} {\bibfnamefont {S.}~\bibnamefont
  {Djurdji\'{c}~Mijin}}, \bibinfo {author} {\bibfnamefont {A.}~\bibnamefont
  {Baum}}, \bibinfo {author} {\bibfnamefont {J.}~\bibnamefont {Bekaert}},
  \bibinfo {author} {\bibfnamefont {A.}~\bibnamefont {\ifmmode \check{S}\else
  \v{S}\fi{}olaji\ifmmode~\acute{c}\else \'{c}\fi{}}}, \bibinfo {author}
  {\bibfnamefont {J.}~\bibnamefont {Pe\ifmmode \check{s}\else
  \v{s}\fi{}i\ifmmode~\acute{c}\else \'{c}\fi{}}}, \bibinfo {author}
  {\bibfnamefont {Y.}~\bibnamefont {Liu}}, \bibinfo {author} {\bibfnamefont
  {G.}~\bibnamefont {He}}, \bibinfo {author} {\bibfnamefont {M.~V.}\
  \bibnamefont {Milo\ifmmode \check{s}\else
  \v{s}\fi{}evi\ifmmode~\acute{c}\else \'{c}\fi{}}}, \bibinfo {author}
  {\bibfnamefont {C.}~\bibnamefont {Petrovic}}, \bibinfo {author}
  {\bibfnamefont {Z.~V.}\ \bibnamefont {Popovi\ifmmode~\acute{c}\else
  \'{c}\fi{}}}, \bibinfo {author} {\bibfnamefont {R.}~\bibnamefont {Hackl}},\
  and\ \bibinfo {author} {\bibfnamefont {N.}~\bibnamefont
  {Lazarevi\ifmmode~\acute{c}\else \'{c}\fi{}}},\ }\bibfield  {title} {\bibinfo
  {title} {{Probing charge density wave phases and the Mott transition in
  1$T$-TaS$_2$ by inelastic light scattering}},\ }\href
  {https://doi.org/10.1103/PhysRevB.103.245133} {\bibfield  {journal} {\bibinfo
   {journal} {Phys. Rev. B}\ }\textbf {\bibinfo {volume} {103}},\ \bibinfo
  {pages} {245133} (\bibinfo {year} {2021})}\BibitemShut {NoStop}%
\bibitem [{\citenamefont {Ramos}\ \emph {et~al.}(2019)\citenamefont {Ramos},
  \citenamefont {Plumadore}, \citenamefont {Boddison-Chouinard}, \citenamefont
  {Hla}, \citenamefont {Guest}, \citenamefont {Gosztola}, \citenamefont
  {Pimenta},\ and\ \citenamefont {Luican-Mayer}}]{Ramos_2019}%
  \BibitemOpen
  \bibfield  {author} {\bibinfo {author} {\bibfnamefont {S.~L. L.~M.}\
  \bibnamefont {Ramos}}, \bibinfo {author} {\bibfnamefont {R.}~\bibnamefont
  {Plumadore}}, \bibinfo {author} {\bibfnamefont {J.}~\bibnamefont
  {Boddison-Chouinard}}, \bibinfo {author} {\bibfnamefont {S.~W.}\ \bibnamefont
  {Hla}}, \bibinfo {author} {\bibfnamefont {J.~R.}\ \bibnamefont {Guest}},
  \bibinfo {author} {\bibfnamefont {D.~J.}\ \bibnamefont {Gosztola}}, \bibinfo
  {author} {\bibfnamefont {M.~A.}\ \bibnamefont {Pimenta}},\ and\ \bibinfo
  {author} {\bibfnamefont {A.}~\bibnamefont {Luican-Mayer}},\ }\bibfield
  {title} {\bibinfo {title} {{Suppression of the commensurate charge density
  wave phase in ultrathin 1$T$-TaS$_2$ evidenced by Raman hyperspectral
  analysis}},\ }\href {https://doi.org/10.1103/PhysRevB.100.165414} {\bibfield
  {journal} {\bibinfo  {journal} {Phys. Rev. B}\ }\textbf {\bibinfo {volume}
  {100}},\ \bibinfo {pages} {165414} (\bibinfo {year} {2019})}\BibitemShut
  {NoStop}%
\bibitem [{\citenamefont {Sanders}\ \emph {et~al.}(2016)\citenamefont
  {Sanders}, \citenamefont {Dendzik}, \citenamefont {Ngankeu}, \citenamefont
  {Eich}, \citenamefont {Bruix}, \citenamefont {Bianchi}, \citenamefont {Miwa},
  \citenamefont {Hammer}, \citenamefont {Khajetoorians},\ and\ \citenamefont
  {Hofmann}}]{Sanders_2016}%
  \BibitemOpen
  \bibfield  {author} {\bibinfo {author} {\bibfnamefont {C.~E.}\ \bibnamefont
  {Sanders}}, \bibinfo {author} {\bibfnamefont {M.}~\bibnamefont {Dendzik}},
  \bibinfo {author} {\bibfnamefont {A.~S.}\ \bibnamefont {Ngankeu}}, \bibinfo
  {author} {\bibfnamefont {A.}~\bibnamefont {Eich}}, \bibinfo {author}
  {\bibfnamefont {A.}~\bibnamefont {Bruix}}, \bibinfo {author} {\bibfnamefont
  {M.}~\bibnamefont {Bianchi}}, \bibinfo {author} {\bibfnamefont {J.~A.}\
  \bibnamefont {Miwa}}, \bibinfo {author} {\bibfnamefont {B.}~\bibnamefont
  {Hammer}}, \bibinfo {author} {\bibfnamefont {A.~A.}\ \bibnamefont
  {Khajetoorians}},\ and\ \bibinfo {author} {\bibfnamefont {P.}~\bibnamefont
  {Hofmann}},\ }\bibfield  {title} {\bibinfo {title} {{Crystalline and
  electronic structure of single-layer 1$T$-TaS$_2$}},\ }\href
  {https://doi.org/10.1103/PhysRevB.94.081404} {\bibfield  {journal} {\bibinfo
  {journal} {Phys. Rev. B}\ }\textbf {\bibinfo {volume} {94}},\ \bibinfo
  {pages} {081404} (\bibinfo {year} {2016})}\BibitemShut {NoStop}%
\bibitem [{\citenamefont {Luo}\ \emph {et~al.}(2021)\citenamefont {Luo},
  \citenamefont {Obeysekera}, \citenamefont {Won}, \citenamefont {Sung},
  \citenamefont {Schnitzer}, \citenamefont {Hovden}, \citenamefont {Cheong},
  \citenamefont {Yang}, \citenamefont {Sun},\ and\ \citenamefont
  {Zhao}}]{Luo_2021}%
  \BibitemOpen
  \bibfield  {author} {\bibinfo {author} {\bibfnamefont {X.}~\bibnamefont
  {Luo}}, \bibinfo {author} {\bibfnamefont {D.}~\bibnamefont {Obeysekera}},
  \bibinfo {author} {\bibfnamefont {C.}~\bibnamefont {Won}}, \bibinfo {author}
  {\bibfnamefont {S.~H.}\ \bibnamefont {Sung}}, \bibinfo {author}
  {\bibfnamefont {N.}~\bibnamefont {Schnitzer}}, \bibinfo {author}
  {\bibfnamefont {R.}~\bibnamefont {Hovden}}, \bibinfo {author} {\bibfnamefont
  {S.-W.}\ \bibnamefont {Cheong}}, \bibinfo {author} {\bibfnamefont
  {J.}~\bibnamefont {Yang}}, \bibinfo {author} {\bibfnamefont {K.}~\bibnamefont
  {Sun}},\ and\ \bibinfo {author} {\bibfnamefont {L.}~\bibnamefont {Zhao}},\
  }\bibfield  {title} {\bibinfo {title} {{Ultrafast modulations and detection
  of a ferro-rotational charge density wave using time-resolved electric
  quadrupole second harmonic generation}},\ }\href
  {https://doi.org/10.1103/PhysRevLett.127.126401} {\bibfield  {journal}
  {\bibinfo  {journal} {Phys. Rev. Lett.}\ }\textbf {\bibinfo {volume} {127}},\
  \bibinfo {pages} {126401} (\bibinfo {year} {2021})}\BibitemShut {NoStop}%
\bibitem [{\citenamefont {Yoshida}\ \emph {et~al.}(2014)\citenamefont
  {Yoshida}, \citenamefont {Zhang}, \citenamefont {Ye}, \citenamefont {Suzuki},
  \citenamefont {Imai}, \citenamefont {Kimura}, \citenamefont {Fujiwara},\ and\
  \citenamefont {Iwasa}}]{Yoshida_2014}%
  \BibitemOpen
  \bibfield  {author} {\bibinfo {author} {\bibfnamefont {M.}~\bibnamefont
  {Yoshida}}, \bibinfo {author} {\bibfnamefont {Y.}~\bibnamefont {Zhang}},
  \bibinfo {author} {\bibfnamefont {J.}~\bibnamefont {Ye}}, \bibinfo {author}
  {\bibfnamefont {R.}~\bibnamefont {Suzuki}}, \bibinfo {author} {\bibfnamefont
  {Y.}~\bibnamefont {Imai}}, \bibinfo {author} {\bibfnamefont {S.}~\bibnamefont
  {Kimura}}, \bibinfo {author} {\bibfnamefont {A.}~\bibnamefont {Fujiwara}},\
  and\ \bibinfo {author} {\bibfnamefont {Y.}~\bibnamefont {Iwasa}},\ }\bibfield
   {title} {\bibinfo {title} {{Controlling charge-density-wave states in
  nano-thick crystals of 1$T$-TaS$_2$}},\ }\href
  {https://doi.org/10.1038/srep07302} {\bibfield  {journal} {\bibinfo
  {journal} {Sci. Rep.}\ }\textbf {\bibinfo {volume} {4}},\ \bibinfo {pages}
  {7302} (\bibinfo {year} {2014})}\BibitemShut {NoStop}%
\bibitem [{\citenamefont {Tsen}\ \emph {et~al.}(2015)\citenamefont {Tsen},
  \citenamefont {Hovden}, \citenamefont {Wang}, \citenamefont {Kim},
  \citenamefont {Okamoto}, \citenamefont {Spoth}, \citenamefont {Liu},
  \citenamefont {Lu}, \citenamefont {Sun}, \citenamefont {Hone}, \citenamefont
  {Kourkoutis}, \citenamefont {Kim},\ and\ \citenamefont
  {Pasupathy}}]{Tsen2015}%
  \BibitemOpen
  \bibfield  {author} {\bibinfo {author} {\bibfnamefont {A.~W.}\ \bibnamefont
  {Tsen}}, \bibinfo {author} {\bibfnamefont {R.}~\bibnamefont {Hovden}},
  \bibinfo {author} {\bibfnamefont {D.}~\bibnamefont {Wang}}, \bibinfo {author}
  {\bibfnamefont {Y.~D.}\ \bibnamefont {Kim}}, \bibinfo {author} {\bibfnamefont
  {J.}~\bibnamefont {Okamoto}}, \bibinfo {author} {\bibfnamefont {K.~A.}\
  \bibnamefont {Spoth}}, \bibinfo {author} {\bibfnamefont {Y.}~\bibnamefont
  {Liu}}, \bibinfo {author} {\bibfnamefont {W.}~\bibnamefont {Lu}}, \bibinfo
  {author} {\bibfnamefont {Y.}~\bibnamefont {Sun}}, \bibinfo {author}
  {\bibfnamefont {J.~C.}\ \bibnamefont {Hone}}, \bibinfo {author}
  {\bibfnamefont {L.~F.}\ \bibnamefont {Kourkoutis}}, \bibinfo {author}
  {\bibfnamefont {P.}~\bibnamefont {Kim}},\ and\ \bibinfo {author}
  {\bibfnamefont {A.~N.}\ \bibnamefont {Pasupathy}},\ }\bibfield  {title}
  {\bibinfo {title} {{Structure and control of charge density waves in
  two-dimensional 1$T$-TaS$_2$}},\ }\href
  {https://doi.org/10.1073/pnas.1512092112} {\bibfield  {journal} {\bibinfo
  {journal} {Proc. Natl. Acad. Sci. U.S.A.}\ }\textbf {\bibinfo {volume}
  {112}},\ \bibinfo {pages} {15054} (\bibinfo {year} {2015})}\BibitemShut
  {NoStop}%
\bibitem [{\citenamefont {Boix-Constant}\ \emph {et~al.}(2021)\citenamefont
  {Boix-Constant}, \citenamefont {Ma{\~n}as-Valero}, \citenamefont
  {C{\'o}rdoba}, \citenamefont {Baldov{\'\i}}, \citenamefont {Rubio},\ and\
  \citenamefont {Coronado}}]{Constant2021}%
  \BibitemOpen
  \bibfield  {author} {\bibinfo {author} {\bibfnamefont {C.}~\bibnamefont
  {Boix-Constant}}, \bibinfo {author} {\bibfnamefont {S.}~\bibnamefont
  {Ma{\~n}as-Valero}}, \bibinfo {author} {\bibfnamefont {R.}~\bibnamefont
  {C{\'o}rdoba}}, \bibinfo {author} {\bibfnamefont {J.~J.}\ \bibnamefont
  {Baldov{\'\i}}}, \bibinfo {author} {\bibfnamefont {{\'A}.}~\bibnamefont
  {Rubio}},\ and\ \bibinfo {author} {\bibfnamefont {E.}~\bibnamefont
  {Coronado}},\ }\bibfield  {title} {\bibinfo {title} {{Out-of-Plane Transport
  of 1$T$-TaS$_2$/Graphene-Based van der Waals Heterostructures}},\ }\href
  {https://doi.org/10.1021/acsnano.1c03012} {\bibfield  {journal} {\bibinfo
  {journal} {ACS Nano}\ }\textbf {\bibinfo {volume} {15}},\ \bibinfo {pages}
  {11898} (\bibinfo {year} {2021})}\BibitemShut {NoStop}%
\bibitem [{\citenamefont {Darancet}\ \emph {et~al.}(2014)\citenamefont
  {Darancet}, \citenamefont {Millis},\ and\ \citenamefont
  {Marianetti}}]{Darancet_2014}%
  \BibitemOpen
  \bibfield  {author} {\bibinfo {author} {\bibfnamefont {P.}~\bibnamefont
  {Darancet}}, \bibinfo {author} {\bibfnamefont {A.~J.}\ \bibnamefont
  {Millis}},\ and\ \bibinfo {author} {\bibfnamefont {C.~A.}\ \bibnamefont
  {Marianetti}},\ }\bibfield  {title} {\bibinfo {title} {{Three-dimensional
  metallic and two-dimensional insulating behavior in octahedral tantalum
  dichalcogenides}},\ }\href {https://doi.org/10.1103/PhysRevB.90.045134}
  {\bibfield  {journal} {\bibinfo  {journal} {Phys. Rev. B}\ }\textbf {\bibinfo
  {volume} {90}},\ \bibinfo {pages} {045134} (\bibinfo {year}
  {2014})}\BibitemShut {NoStop}%
\bibitem [{\citenamefont {Bae}\ \emph {et~al.}(2025)\citenamefont {Bae},
  \citenamefont {Valentí}, \citenamefont {Mazin},\ and\ \citenamefont
  {Yan}}]{Bae2025}%
  \BibitemOpen
  \bibfield  {author} {\bibinfo {author} {\bibfnamefont {H.}~\bibnamefont
  {Bae}}, \bibinfo {author} {\bibfnamefont {R.}~\bibnamefont {Valentí}},
  \bibinfo {author} {\bibfnamefont {I.~I.}\ \bibnamefont {Mazin}},\ and\
  \bibinfo {author} {\bibfnamefont {B.}~\bibnamefont {Yan}},\ }\bibfield
  {title} {\bibinfo {title} {Designing flat bands, localized and itinerant
  states in {TaS$_2$} trilayer heterostructures},\ }\href
  {https://doi.org/10.1038/s41535-025-00812-0} {\bibfield  {journal} {\bibinfo
  {journal} {npj Quantum Mater.}\ }\textbf {\bibinfo {volume} {10}},\ \bibinfo
  {pages} {92} (\bibinfo {year} {2025})}\BibitemShut {NoStop}%
\bibitem [{\citenamefont {Butler}\ \emph {et~al.}(2020)\citenamefont {Butler},
  \citenamefont {Yoshida}, \citenamefont {Hanaguri},\ and\ \citenamefont
  {Iwasa}}]{Butler2020}%
  \BibitemOpen
  \bibfield  {author} {\bibinfo {author} {\bibfnamefont {C.~J.}\ \bibnamefont
  {Butler}}, \bibinfo {author} {\bibfnamefont {M.}~\bibnamefont {Yoshida}},
  \bibinfo {author} {\bibfnamefont {T.}~\bibnamefont {Hanaguri}},\ and\
  \bibinfo {author} {\bibfnamefont {Y.}~\bibnamefont {Iwasa}},\ }\bibfield
  {title} {\bibinfo {title} {Mottness versus unit-cell doubling as the driver
  of the insulating state in {1$T$-TaS}$_2$},\ }\href
  {https://doi.org/10.1038/s41467-020-16132-9} {\bibfield  {journal} {\bibinfo
  {journal} {Nat. Commun.}\ }\textbf {\bibinfo {volume} {11}},\ \bibinfo
  {pages} {2477} (\bibinfo {year} {2020})}\BibitemShut {NoStop}%
\bibitem [{\citenamefont {Wu}\ \emph {et~al.}(2022)\citenamefont {Wu},
  \citenamefont {Bu}, \citenamefont {Zhang}, \citenamefont {Fei}, \citenamefont
  {Zheng}, \citenamefont {Gao}, \citenamefont {Luo}, \citenamefont {Liu},
  \citenamefont {Sun},\ and\ \citenamefont {Yin}}]{Wu2022}%
  \BibitemOpen
  \bibfield  {author} {\bibinfo {author} {\bibfnamefont {Z.}~\bibnamefont
  {Wu}}, \bibinfo {author} {\bibfnamefont {K.}~\bibnamefont {Bu}}, \bibinfo
  {author} {\bibfnamefont {W.}~\bibnamefont {Zhang}}, \bibinfo {author}
  {\bibfnamefont {Y.}~\bibnamefont {Fei}}, \bibinfo {author} {\bibfnamefont
  {Y.}~\bibnamefont {Zheng}}, \bibinfo {author} {\bibfnamefont
  {J.}~\bibnamefont {Gao}}, \bibinfo {author} {\bibfnamefont {X.}~\bibnamefont
  {Luo}}, \bibinfo {author} {\bibfnamefont {Z.}~\bibnamefont {Liu}}, \bibinfo
  {author} {\bibfnamefont {Y.-P.}\ \bibnamefont {Sun}},\ and\ \bibinfo {author}
  {\bibfnamefont {Y.}~\bibnamefont {Yin}},\ }\bibfield  {title} {\bibinfo
  {title} {Effect of stacking order on the electronic state of
  {1$T$-TaS$_2$}},\ }\href {https://doi.org/10.1103/PhysRevB.105.035109}
  {\bibfield  {journal} {\bibinfo  {journal} {Phys. Rev. B}\ }\textbf {\bibinfo
  {volume} {105}},\ \bibinfo {pages} {035109} (\bibinfo {year}
  {2022})}\BibitemShut {NoStop}%
\bibitem [{\citenamefont {{N. F. Mott}}(1968)}]{Mott_1968}%
  \BibitemOpen
  \bibfield  {author} {\bibinfo {author} {\bibnamefont {{N. F. Mott}}},\
  }\bibfield  {title} {\bibinfo {title} {Conduction in glasses containing
  transition metal ions},\ }\href
  {https://doi.org/https://doi.org/10.1016/0022-3093(68)90002-1} {\bibfield
  {journal} {\bibinfo  {journal} {J. Non-Cryst. Solids}\ }\textbf {\bibinfo
  {volume} {1}},\ \bibinfo {pages} {1} (\bibinfo {year} {1968})}\BibitemShut
  {NoStop}%
\bibitem [{\citenamefont {Efros}\ and\ \citenamefont
  {Shklovskii}(1975)}]{Efros_1975}%
  \BibitemOpen
  \bibfield  {author} {\bibinfo {author} {\bibfnamefont {A.~L.}\ \bibnamefont
  {Efros}}\ and\ \bibinfo {author} {\bibfnamefont {B.~I.}\ \bibnamefont
  {Shklovskii}},\ }\bibfield  {title} {\bibinfo {title} {{Coulomb gap and low
  temperature conductivity of disordered systems}},\ }\href
  {https://doi.org/10.1088/0022-3719/8/4/003} {\bibfield  {journal} {\bibinfo
  {journal} {J. Phys. C.}\ }\textbf {\bibinfo {volume} {8}},\ \bibinfo {pages}
  {L49} (\bibinfo {year} {1975})}\BibitemShut {NoStop}%
\bibitem [{\citenamefont {Inada}\ \emph {et~al.}(1983)\citenamefont {Inada},
  \citenamefont {Ōnuki},\ and\ \citenamefont {Tanuma}}]{Inada1983}%
  \BibitemOpen
  \bibfield  {author} {\bibinfo {author} {\bibfnamefont {R.}~\bibnamefont
  {Inada}}, \bibinfo {author} {\bibfnamefont {Y.}~\bibnamefont {Ōnuki}},\ and\
  \bibinfo {author} {\bibfnamefont {S.-i.}\ \bibnamefont {Tanuma}},\ }\bibfield
   {title} {\bibinfo {title} {Anderson localization and crystalline defects of
  {1$T$-TaS$_2$}},\ }\href {https://doi.org/10.1143/JPSJ.52.3536} {\bibfield
  {journal} {\bibinfo  {journal} {J. Phys. Soc. Jpn.}\ }\textbf {\bibinfo
  {volume} {52}},\ \bibinfo {pages} {3536} (\bibinfo {year}
  {1983})}\BibitemShut {NoStop}%
\bibitem [{\citenamefont {{A. L. Efros, N. Van Lien, and B. I.
  Shklovskii}}(1979)}]{Efros_1979}%
  \BibitemOpen
  \bibfield  {author} {\bibinfo {author} {\bibnamefont {{A. L. Efros, N. Van
  Lien, and B. I. Shklovskii}}},\ }\bibfield  {title} {\bibinfo {title}
  {Variable range hopping in doped crystalline semiconductors},\ }\href
  {https://doi.org/https://doi.org/10.1016/0038-1098(79)90484-8} {\bibfield
  {journal} {\bibinfo  {journal} {Solid State Commun.}\ }\textbf {\bibinfo
  {volume} {32}},\ \bibinfo {pages} {851} (\bibinfo {year} {1979})}\BibitemShut
  {NoStop}%
\bibitem [{\citenamefont {Bao}\ \emph {et~al.}(2022)\citenamefont {Bao},
  \citenamefont {Yang},\ and\ \citenamefont {Wang}}]{Bao_2022}%
  \BibitemOpen
  \bibfield  {author} {\bibinfo {author} {\bibfnamefont {J.}~\bibnamefont
  {Bao}}, \bibinfo {author} {\bibfnamefont {L.}~\bibnamefont {Yang}},\ and\
  \bibinfo {author} {\bibfnamefont {D.}~\bibnamefont {Wang}},\ }\bibfield
  {title} {\bibinfo {title} {{Influence of torsional deformation on the
  electronic structure and optical properties of 1$T$-TaS$_2$ monolayer}},\
  }\href {https://doi.org/https://doi.org/10.1016/j.molstruc.2022.132667}
  {\bibfield  {journal} {\bibinfo  {journal} {J. Mol. Struct}\ }\textbf
  {\bibinfo {volume} {1258}},\ \bibinfo {pages} {132667} (\bibinfo {year}
  {2022})}\BibitemShut {NoStop}%
\bibitem [{\citenamefont {Zhang}\ \emph {et~al.}(2024)\citenamefont {Zhang},
  \citenamefont {Yan},\ and\ \citenamefont {Li}}]{Zhang2024}%
  \BibitemOpen
  \bibfield  {author} {\bibinfo {author} {\bibfnamefont {X.}~\bibnamefont
  {Zhang}}, \bibinfo {author} {\bibfnamefont {S.}~\bibnamefont {Yan}},\ and\
  \bibinfo {author} {\bibfnamefont {G.}~\bibnamefont {Li}},\ }\bibfield
  {title} {\bibinfo {title} {{Enhanced charge density wave and the cluster Mott
  state driven by nonlocal electronic correlations in 1$T$-NbS$_{2}$}},\ }\href
  {https://doi.org/10.1103/PhysRevB.110.235137} {\bibfield  {journal} {\bibinfo
   {journal} {Phys. Rev. B}\ }\textbf {\bibinfo {volume} {110}},\ \bibinfo
  {pages} {235137} (\bibinfo {year} {2024})}\BibitemShut {NoStop}%
\bibitem [{\citenamefont {Hasaien}\ \emph {et~al.}(2025)\citenamefont
  {Hasaien}, \citenamefont {Wu}, \citenamefont {Shi}, \citenamefont {Zhai},
  \citenamefont {Wu}, \citenamefont {Liu}, \citenamefont {Zhou}, \citenamefont
  {Chen},\ and\ \citenamefont {Zhao}}]{Hasaien2025}%
  \BibitemOpen
  \bibfield  {author} {\bibinfo {author} {\bibfnamefont {J.}~\bibnamefont
  {Hasaien}}, \bibinfo {author} {\bibfnamefont {Y.}~\bibnamefont {Wu}},
  \bibinfo {author} {\bibfnamefont {M.}~\bibnamefont {Shi}}, \bibinfo {author}
  {\bibfnamefont {Y.}~\bibnamefont {Zhai}}, \bibinfo {author} {\bibfnamefont
  {Q.}~\bibnamefont {Wu}}, \bibinfo {author} {\bibfnamefont {Z.}~\bibnamefont
  {Liu}}, \bibinfo {author} {\bibfnamefont {Y.}~\bibnamefont {Zhou}}, \bibinfo
  {author} {\bibfnamefont {X.}~\bibnamefont {Chen}},\ and\ \bibinfo {author}
  {\bibfnamefont {J.}~\bibnamefont {Zhao}},\ }\bibfield  {title} {\bibinfo
  {title} {Emergent quantum state unveiled by ultrafast collective dynamics in
  {1$T$-TaS$_2$}},\ }\href {https://doi.org/10.1073/pnas.2406464122} {\bibfield
   {journal} {\bibinfo  {journal} {Proc. Natl. Acad. Sci. U.S.A.}\ }\textbf
  {\bibinfo {volume} {122}},\ \bibinfo {pages} {e2406464122} (\bibinfo {year}
  {2025})}\BibitemShut {NoStop}%
\bibitem [{\citenamefont {{Y. Chen et al.,}}(2020)}]{Chen_2020}%
  \BibitemOpen
  \bibfield  {author} {\bibinfo {author} {\bibnamefont {{Y. Chen et al.,}}},\
  }\bibfield  {title} {\bibinfo {title} {{Strong correlations and orbital
  texture in single-layer 1$T$-TaSe$_2$}},\ }\href
  {https://doi.org/10.1038/s41567-019-0744-9} {\bibfield  {journal} {\bibinfo
  {journal} {Nat. Phys.}\ }\textbf {\bibinfo {volume} {16}},\ \bibinfo {pages}
  {218} (\bibinfo {year} {2020})}\BibitemShut {NoStop}%
\bibitem [{\citenamefont {Jiang}\ \emph {et~al.}(2021)\citenamefont {Jiang},
  \citenamefont {Hu}, \citenamefont {Zhao}, \citenamefont {Li}, \citenamefont
  {Xu}, \citenamefont {Liu}, \citenamefont {Cui},\ and\ \citenamefont
  {Ren}}]{Jiang_2021}%
  \BibitemOpen
  \bibfield  {author} {\bibinfo {author} {\bibfnamefont {T.}~\bibnamefont
  {Jiang}}, \bibinfo {author} {\bibfnamefont {T.}~\bibnamefont {Hu}}, \bibinfo
  {author} {\bibfnamefont {G.-D.}\ \bibnamefont {Zhao}}, \bibinfo {author}
  {\bibfnamefont {Y.}~\bibnamefont {Li}}, \bibinfo {author} {\bibfnamefont
  {S.}~\bibnamefont {Xu}}, \bibinfo {author} {\bibfnamefont {C.}~\bibnamefont
  {Liu}}, \bibinfo {author} {\bibfnamefont {Y.}~\bibnamefont {Cui}},\ and\
  \bibinfo {author} {\bibfnamefont {W.}~\bibnamefont {Ren}},\ }\bibfield
  {title} {\bibinfo {title} {{Two-dimensional charge density waves in
  $\mathrm{Ta}{X}_{2}$ ($X=\mathrm{S}$, Se, Te) from first principles}},\
  }\href {https://doi.org/10.1103/PhysRevB.104.075147} {\bibfield  {journal}
  {\bibinfo  {journal} {Phys. Rev. B}\ }\textbf {\bibinfo {volume} {104}},\
  \bibinfo {pages} {075147} (\bibinfo {year} {2021})}\BibitemShut {NoStop}%
\bibitem [{\citenamefont {Mañas-Valero}\ \emph {et~al.}(2021)\citenamefont
  {Mañas-Valero}, \citenamefont {{B. M. Huddart}}, \citenamefont {Lancaster}
  \emph {et~al.}}]{ManasValero2021}%
  \BibitemOpen
  \bibfield  {author} {\bibinfo {author} {\bibfnamefont {S.}~\bibnamefont
  {Mañas-Valero}}, \bibinfo {author} {\bibnamefont {{B. M. Huddart}}},
  \bibinfo {author} {\bibfnamefont {T.}~\bibnamefont {Lancaster}}, \emph
  {et~al.},\ }\bibfield  {title} {\bibinfo {title} {{Quantum phases and spin
  liquid properties of 1$T$-TaS$_2$}},\ }\href
  {https://doi.org/10.1038/s41535-021-00367-w} {\bibfield  {journal} {\bibinfo
  {journal} {npj Quantum Mater}\ }\textbf {\bibinfo {volume} {6}},\ \bibinfo
  {pages} {69} (\bibinfo {year} {2021})}\BibitemShut {NoStop}%
\bibitem [{\citenamefont {Liu}\ \emph {et~al.}(2026)\citenamefont {Liu},
  \citenamefont {Liu}, \citenamefont {Luo}, \citenamefont {Huang},
  \citenamefont {Watanabe}, \citenamefont {Taniguchi}, \citenamefont {Wang},
  \citenamefont {Wen}, \citenamefont {Lu},\ and\ \citenamefont {Xi}}]{DAS}%
  \BibitemOpen
  \bibfield  {author} {\bibinfo {author} {\bibfnamefont {G.}~\bibnamefont
  {Liu}}, \bibinfo {author} {\bibfnamefont {Y.}~\bibnamefont {Liu}}, \bibinfo
  {author} {\bibfnamefont {Q.}~\bibnamefont {Luo}}, \bibinfo {author}
  {\bibfnamefont {Z.}~\bibnamefont {Huang}}, \bibinfo {author} {\bibfnamefont
  {K.}~\bibnamefont {Watanabe}}, \bibinfo {author} {\bibfnamefont
  {T.}~\bibnamefont {Taniguchi}}, \bibinfo {author} {\bibfnamefont
  {M.}~\bibnamefont {Wang}}, \bibinfo {author} {\bibfnamefont {J.}~\bibnamefont
  {Wen}}, \bibinfo {author} {\bibfnamefont {Y.}~\bibnamefont {Lu}},\ and\
  \bibinfo {author} {\bibfnamefont {X.}~\bibnamefont {Xi}},\ }\href
  {https://doi.org/10.6084/m9.figshare.32731371} {\bibinfo {title} {Data for
  ``{Strongly Enhanced Charge-Density Waves and Correlated Insulating State in
  Atomically Thin 1$T$-TaS$_2$}''}} (\bibinfo {year} {2026})\BibitemShut
  {NoStop}%
\end{thebibliography}
\end{document}